\documentclass[a4paper,11pt]{article}
\pdfoutput=1 
\date{}
\usepackage{jheppub}
\usepackage{amsmath,amssymb,amsfonts,amsxtra, mathrsfs,graphics,graphicx,amsthm,epsfig, youngtab,bm,longtable,float,tikz,empheq}
\usepackage[margin=0.5cm]{caption}
\allowdisplaybreaks[1]
\usetikzlibrary{positioning}
\newcommand{\ba}{\begin{array}}
\newcommand{\ea}{\end{array}}
\newcommand{\bi}{\begin{itemize}}
\newcommand{\ei}{\end{itemize}}
\def\vec#1{\bm{#1}}
\def\bea#1\eea{\allowdisplaybreaks \begin{align}#1\end{align}}
 \newcommand{\ben}{\begin{enumerate}}
\newcommand{\een}{\end{enumerate}}
\newcommand{\bean}{\begin{eqnarray*}}
\newcommand{\eean}{\end{eqnarray*}}
\newcommand{\eref}[1]{(\ref{#1})}

\newcommand{\nn}{\nonumber}

\newcommand{\tr}{\mathrm{Tr}}

\newcommand{\tq}{\widetilde{q}}
\newcommand{\tE}{\widetilde{E}}
\newcommand{\tx}{\widetilde{x}}

\newcommand{\BC}{\mathbb{C}}

\newcommand{\BZ}{\mathbb{Z}}

\newcommand{\comment}[1]{}

\newcommand{\CN}{{\cal N}}

\newcommand{\CH}{{\cal H}}
\newcommand{\CZ}{{\cal Z}}

\newcommand{\CI}{{\cal I}}
\newcommand{\diag}{\mathrm{diag}}

\newcommand{\ie}{{\it i.e.}}

\newcommand{\tti}{\widetilde{t}}

\newcommand{\cN}{\mathcal{N}}

\newcommand{\sh}{\sinh \pi}
\newcommand{\ch}{\cosh \pi}

\newcommand{\Secref}[1]{Section~\ref{#1}}
\newcommand{\secref}[1]{Sec.~\ref{#1}}
\newcommand{\Appref}[1]{Appendix~\ref{#1}}
\newcommand{\appref}[1]{App.~\ref{#1}}

\newcommand{\figref}[1]{Fig.~\ref{#1}}
\renewcommand{\eqref}[1]{(\ref{#1})}

\title{On Three-Dimensional Quiver Gauge Theories of Type B}

\author[a]{Anindya Dey,}
\author[b]{Amihay Hanany,}
\author[c]{Peter Koroteev,}
\author[d,e]{and Noppadol Mekareeya}
\affiliation[a]{New High Energy Theory Center, Rutgers University, Piscataway, NJ}
\affiliation[b]{Theoretical Physics Group, The Blackett Laboratory, \\Imperial College London London, United Kingdom}
\affiliation[c]{Perimeter Institute for Theoretical Physics \\Waterloo, ON}
\affiliation[c]{Department of Mathematics, University of California \\ Davis, CA}
\affiliation[d]{Dipartimento di Fisica, Universit\`a di Milano-Bicocca, \\ Piazza della Scienza 3, I-20126 Milano, Italy}
\affiliation[e]{INFN, sezione di Milano-Bicocca, I-20126 Milano, Italy}
\emailAdd{anindya@physics.utexas.edu}
\emailAdd{a.hanany@imperial.ac.uk}
\emailAdd{pkoroteev@perimeterinstitute.ca}
\emailAdd{n.mekareeya@gmail.com}

\abstract{We study three-dimensional supersymmetric quiver gauge theories with a non-simply laced global symmetry primarily focusing on framed affine $B_{N}$ quiver theories. Using a supersymmetric partition function on a three sphere, and its transformation under S-duality, we study the three-dimensional ADHM quiver for $SO(2N+1)$ instantons with a half-integer Chern-Simons coupling. The theory after S-duality has no Lagrangian, and can not be represented by a single quiver, however its partition function can be conveniently described by a collection of framed affine ${B}_{N}$ quivers. This correspondence can be conjectured to generalize three-dimensional mirror symmetry to theories with nontrivial Chern-Simons terms. In addition, we propose a formula for the superconformal index of a theory described by a framed affine $B_N$ quiver.
}

\preprint{Imperial/TP/16/AH/06}

\begin{document}
\maketitle

\newpage

\section{Introduction and Main Results}\label{Sec:Intro}
String theory and supersymmetric gauge theories have proved to be useful in the study of moduli spaces of Yang-Mills instantons. One of the earliest successes was to give a simple string theory realization \cite{Douglas:xy, Witten:1995gx} of the Atyah-Drinfied-Hitchin-Manin (ADHM) construction \cite{Atiyah:1978ri} for the moduli spaces of instantons for classical gauge groups. As a result of such a string theory construction, these moduli spaces can be identified as the Higgs branches of  supersymmetric quiver gauge theories with eight supercharges; the latter are often referred to as the ADHM quivers. In particular, the ADHM quiver for $k$ $SU(N)$ instantons on $\BC^2$ can be realized on the worldvolume of $k$ D$p$ branes inside the worldvolume of $N$ coincident D$(p+4)$ branes. Similarly for $SO(2N)$, $SO(2N+1)$ or $Sp(N)$ instantons on $\BC^2$, the corresponding ADHM quivers can be described by introducing an appropriate orientifold plane to the aforementioned brane system.  In this paper, we focus mainly on three spacetime dimensions and the corresponding ADHM quiver can be realized from such a brane system with $p=2$.  It should be emphasized that the ADHM quiver theories are available only for instantons for Yang-Mills theories with classical gauge groups.  For the exceptional gauge groups of $E$ type, it turns out that the field theory whose Higgs branch is isomorphic to the corresponding moduli space of instantons can be realized as a circle compactification of the worldvolume theory of M5-branes wrapping Riemann surfaces with appropriate punctures \cite{Gaiotto:2009we, Benini:2009gi, Gaiotto:2012uq, Kimura:2016aa} (also known as 3d Sicilian theories \cite{Benini:2010uu}).  Nevertheless the Lagrangian descriptions of such theories is not known and the generalization of such a construction to the cases of $F_4$ and $G_2$ are not available.
   
In three dimensions, it was found also that the Coulomb branch of certain supersymmetric field theories with eight supercharges (namely, $\CN=4$ supersymmetry) describes the moduli space of instantons.  As was pointed out by \cite{Intriligator:1996ex, deBoer:1996mp, Porrati:1996xi}, the moduli space of $G$-instantons, for $G$ being a simply-laced group $(ADE)$, can be realized as the Coulomb branch of the quiver given by a framed affine Dynkin diagram of group $G$, \ie~ the affine Dynkin diagram with one flavour node attached to the affine gauge node.  (For convenience, this will be denoted by the shorthand notation $[\widehat{G}]$ in the following.)  In particular, for $G$ being of $A$ or $D$ type, such quivers can be obtained by applying three dimensional mirror symmetry \cite{Intriligator:1996ex, Hanany:1996ie, Kapustin:1998fa} to the ADHM quivers associated with $SU(N)$ and $SO(2N)$ instantons on $\BC^2$.   In these cases, Type IIB brane configurations \cite{Hanany:1996ie, Kapustin:1998fa} along with the S-duality provide a convenient way to study quiver descriptions of such field theories.  For $G$ of $E$-type, the corresponding framed affine Dynkin diagrams are precisely the three dimensional mirror theories \cite{Benini:2010uu} of the aforementioned Sicilian theories.  Indeed, the generating function of the holomorphic functions on the Coulomb branch, also known as the Coulomb branch Hilbert series, for the former has been computed \cite{Cremonesi:2013lqa, Cremonesi:2014kwa, Cremonesi:2014vla, Cremonesi:2014xha} and it is in agreement with the result obtained from the Higgs branch of the theories that describe the same moduli space of instantons \cite{Benvenuti:2010pq, Gadde:2011uv, Gaiotto:2012uq, Hanany:2012dm}.
   
One can now generalize the above results to non-simply laced groups $G$. The corresponding affine Dynkin diagrams contain double or triple arrows, whose weakly coupled Lagrangian description is not known to date. In \cite{Cremonesi:2014xha}, a prescription for computing the Coulomb branch Hilbert series for non-simply laced quivers was proposed.  For $G$ being of $B$ and $C$ types the Coulomb branch Hilbert series are in perfect agreement with Higgs branch Hilbert series computed for the $SO(2N+1)$ and $Sp(N)$ ADHM quivers. When $G$ is $F_4$ and $G_2$, the above results passed a number of non-trivial tests. Yet the Lagrangian description of such non-simply laced quivers remains an open question. 

The main goal of the current paper is to gain a better understanding on the physics of three-dimensional quiver theories whose global symmetry is a non-simply laced group. To achieve this goal, we utilize supersymmetric observables including the partition function on round three-sphere \cite{Kapustin:2009kz, Kapustin:2010xq, Benvenuti:2011ga} and the superconformal index \cite{Kim:2009wb, Imamura:2011su, Krattenthaler:2011da, Kapustin:2011jm} to study such field theories. For concreteness, we focus on the case of one $SO(2N+1)$ instanton. The ADHM quiver consists of gauge group $Sp(1)$ with $2N+1$ fundamental half-hypermultipelts. Since the number of half-hypermultipelts is odd,  the theory also has half-integer Chern-Simons coupling at the quantum level due to parity anomaly. 

Supersymmetric partition functions proved to be very effective in studying three dimensional mirror symmetry including the examples which involve non-Lagrangian theories on one side (e.g. circle compactifications of class S theories, see \cite{Dey:2014jk} for details and recent review). In particular one can translate the action of S-duality to the matrix integrals which are used in the expressions for partition functions \cite{Dey:2013nf,Dey:2011pt} and derive the partition function for the mirror dual. This can be used both for verifying the conjectured mirror dualities as well as finding new mirror dual pairs \cite{Dey:2014jk}. In the references above the methods were used for quiver theories with $\CN=4$ supersymmetry, however it appears that the same techniques can be extended to lower supersymmetry, and the current work extends the S-duality transformation to non-simply laced quivers which have $\CN=3$ supersymmetry. The result, however, is partially successful since the expression for the partition function is easy to get, but the mirror theory is much harder to read off from this expression. One does not expect a Lagrangian but it is possible to encode the partition function data in terms of quivers.

We start with the computation of $S^3$ partition function of the ADHM quiver for $k$ $SO(2N+1)$ instantons, namely $Sp(k)$ gauge theory with $2N+1$ flavours of the fundamental hypermultiplets and one anti-symmetric hypermultiplet, with half-integer Chern-Simons coupling for the $Sp(k)$ gauge group, then implement an S-duality transformation on the partition function. The result can be arranged in such a way that the structure of the framed affine ${B}_{N}$ Dynkin diagram becomes apparent.  In particular, the contribution of the double lace in such a Dynkin diagram can be explicitly spelt out from the result. The outcome of such an S-duality transformation also allows us to conjecture the expressions of the partition function on $S^2 \times S^1$.  Upon setting the Chern-Simons level of the original ADHM theory to zero (and the theory is thus parity anomalous), the Coulomb branch limit of the latter partition function reproduces the Hilbert series of one $SO(2N+1)$ instanton on $\mathbb{C}^2$ \cite{Benvenuti:2010pq, Hanany:2012dm, Cremonesi:2014xha}.
   
The paper is organized as follows.  The remainder of this section reviews the ADHM quiver for $SO(2N+1)$ instantons and the main results are stated in \Secref{main}. \Secref{sec:BNcomputation} deals with the $S^3$ partition function for the ADHM quiver for $SO(2N+1)$ instantons with Chern-Simons level $1/2$ and its S-duality transformation. In Sections \ref{sec:SupConfIndex4d} and \Secref{Sec:N2Index}, we state our conjectures for the partition function on $S^2 \times S^1$.

The paper has several appendices. \Appref{Sec:CauchyFourier} describes how to apply Cauchy transform in order to S-dualize the partition function in question. \Appref{YZ} and \Appref{Sec:Znsl} contain technical details of the main computation. \Appref{Sec:GenIndex} contains summary of superconformal indices for Lens spaces. In \Appref{Sec:Folding} we obtain the framed affine $B_3$ quiver by folding the framed affine $D_4$ quiver and analyze its physics using the space of supersymmetric vacua. Finally in \Appref{Sec:HSApp} we discuss the action of folding on the Hilbert series.

\subsection{Double Arrow and Dimension Counting}
\begin{figure}[htbp]
\begin{center}
\includegraphics[scale=0.4]{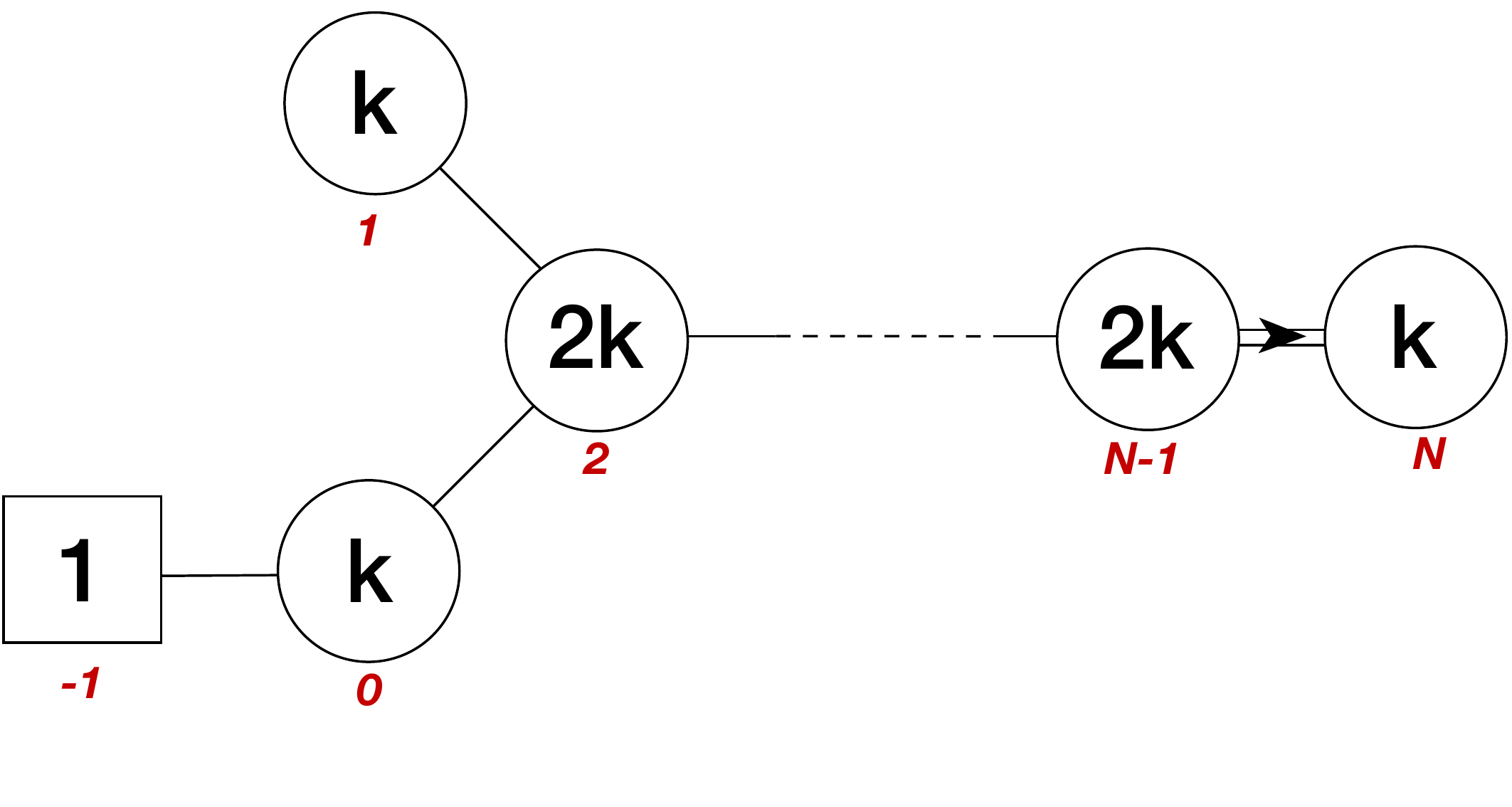}
\caption{The framed affine $B_{N}$ quiver, also denoted by $[\widehat{B}_{N}]$, with ranks of the unitary groups written black and node labels in written red.}
\label{fig:quiverBN}
\end{center}
\end{figure}

The new ingredient of $B_N$-type quivers, which is not present in the $A$ and $D$-type constructions, is the presence of the double arrow which connects the two right-most nodes of the quiver (see \figref{fig:quiverBN}). 
As mentioned in the introduction, this work studies a gauge theory described by a framed affine $B_N$ quiver; therefore we need to understand what kind of `matter' does the double arrow represents. Naively one may try to interpret this `matter' as a bifundamental multiplet of some sort which is charged under the gauge groups corresponding to the nodes at its ends ($N-1$ and $N$ in \figref{fig:quiverBN}). However, as we shall see momentarily, this naive guess fails.

\begin{figure}[!h]
\begin{center}
\includegraphics[scale=0.35]{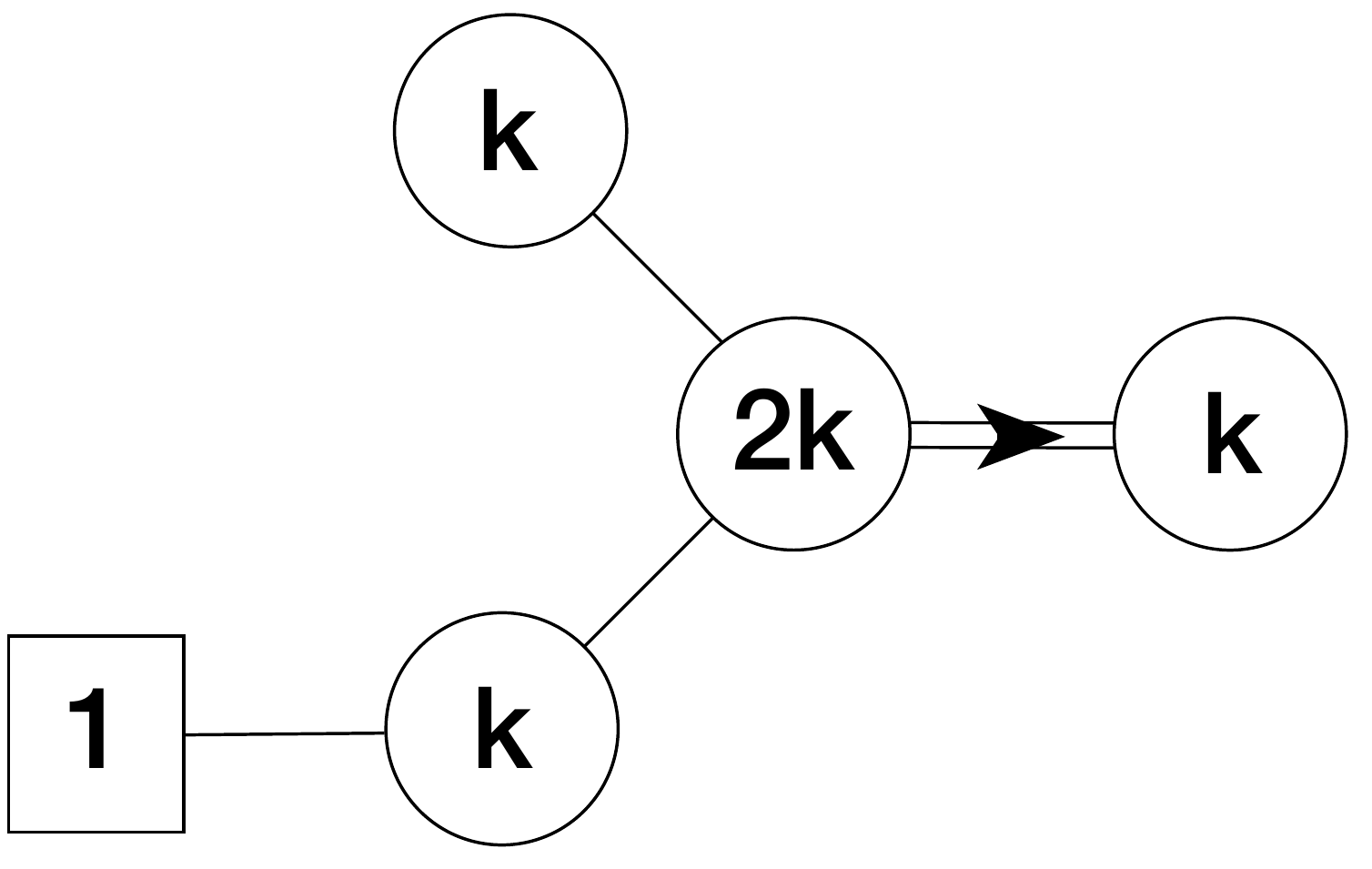}\quad\quad\quad\quad\quad \includegraphics[scale=0.5]{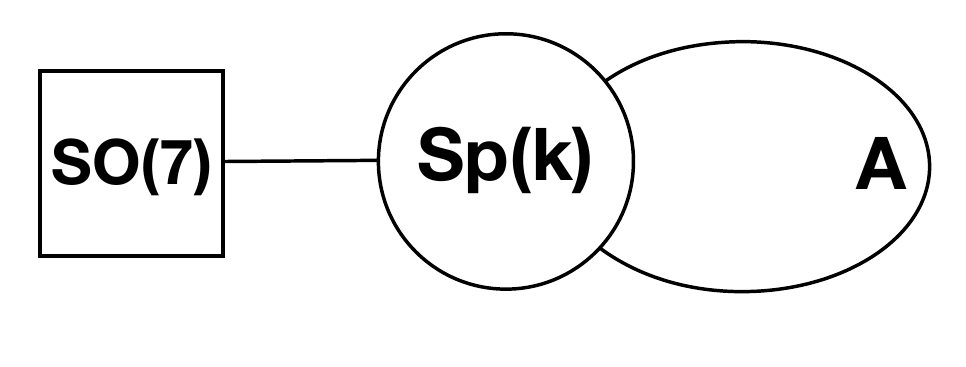}
\caption{ADHM quiver for $k$ $SO(7)$ instantons (right) and its dual quiver with $\widehat B_3$ symmetry (left).}
\label{fig:B4S07Mirror}
\end{center}
\end{figure}

\begin{figure}[!h]
\begin{center}
\includegraphics[scale=0.35]{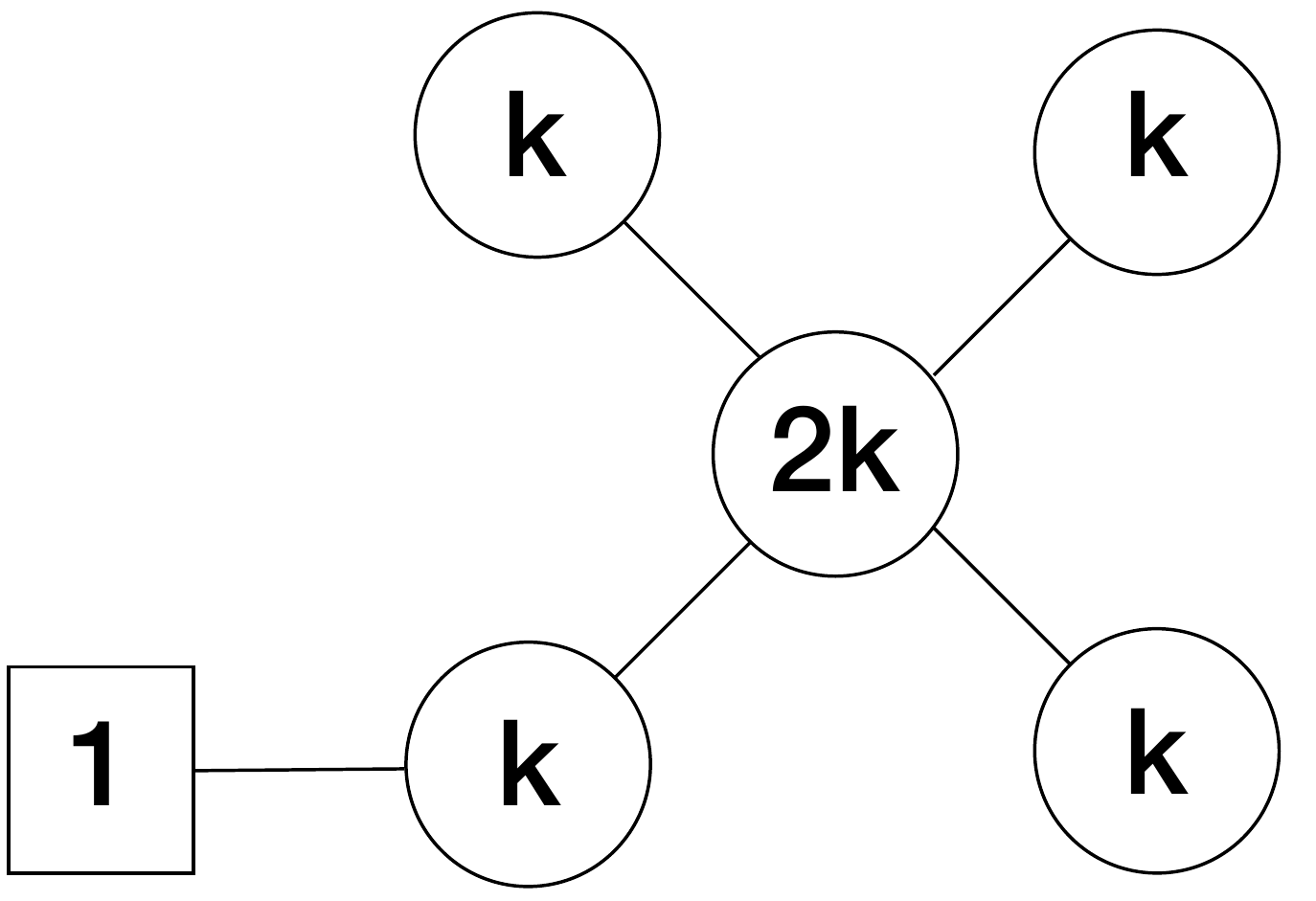}\quad\quad\quad\quad\quad \includegraphics[scale=0.5]{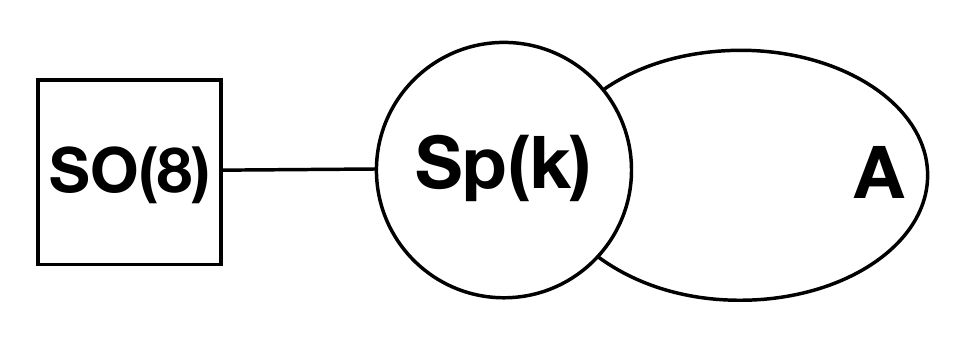}
\caption{ADHM quiver for $k$ $SO(8)$ instantons (right) and its dual quiver with $\widehat D_4$ symmetry (left).}
\label{fig:D4S08Mirror}
\end{center}
\end{figure}

Let us consider $Sp(k)$ theory with 7 fundamental half hypermultiplets, one hypermultiplet in the anti-symmetric representation of $Sp(k)$ and an $SO(7)$ global symmetry. The quiver of its three-dimensional mirror theory can be derived from the S-dual brane construction with orientifold planes \cite{Cremonesi:2014xha} and represents an framed affine $B_3$ Dynkin diagram, see \figref{fig:B4S07Mirror}. 
This quiver can be compared with the quiver for $SO(8)$ global symmetry, which has a mirror quiver that is simply laced (\figref{fig:D4S08Mirror})\footnote{`Folding' of quivers are discussed in \appref{Sec:Folding}}.


Let us compute the quaternionic dimensions of the Coulomb branch of the theory on the right and the dimension of the Higgs branch of the theory on the left and in \figref{fig:B4S07Mirror}. Since the consistency of the theory on the right requires Chern-Simons action with half-integer level, its Coulomb branch is lifted. However, we can still consider the classical Coulomb branch, whose dimension is equal to $k$. From the anticipated duality with the theory on the right of \figref{fig:B4S07Mirror} we expect the dimension of the Higgs branch of the $B_N$ theory to be equal to $k$. At the moment we do not know the contribution of the double arrow to the dimension formula, so we should leave it for a moment as unknown and later derive it from the condition that the quaternionic dimension of the branch should be $k$. One has for the Higgs branch dimension, which is the total number of hypermultipelts minus the number of vector multiplets\footnote{In other words $$\text{dim}_{\text{Higgs}} =\sum_{s(I),t(J)} N^{(I)}_f N^{(J)}_c - \sum_I (N^{I}_c)^2,$$ where the first sum goes over all possible source $s(I)$ and target $t(J)$ nodes of the quiver.} 
\begin{equation}
\text{dim}_{\text{Higgs}} = (k+2k^2+2k^2+2\alpha k^2)-(k^2+(2k)^2+k^2+\beta k^2) = k+(2\alpha-\beta-2)k^2\,,
\end{equation}
which imposes the constraint $2\alpha-\beta-2=0$. For the $D_4$ theory of \figref{fig:D4S08Mirror} the choice is $\alpha=\beta=2$, whereas for the theories in \figref{fig:B4S07Mirror} it is impossible to make both $\alpha$ and $\beta$ integral. Perhaps, the most logical choice is to assign $\alpha =3/2\,,\beta =1$ to account for the single $U(k)$ group left after folding. In any case, the matter sector corresponding to the double arrow in the dual theory with $\widehat{B}_3$ symmetry on its Coulomb brach appears to be non-Lagrangian.
Nevertheless we shall be able to compute the partition function of the framed affine $B_3$ theory and successfully identify the contribution of the matter fields corresponding to the double arrow.

\subsection{Main Results}\label{main}
\begin{itemize}
\item \textbf{Dual of an $\CN=3$ CS-YM theory with symplectic gauge group} \\
We compute the partition function of the dual to the three dimensional supersymmetric $Sp(k)$ Yang-Mills Chern-Simons theory with $2N+1$ $(N \in \BZ)$ half-hypermultiplets, a single antisymmetric hypermultiplet (a singlet for $k=1$) and Chern-Simons level $\kappa \in \BZ/2$. Starting from the partition function of the aforementioned theory on a round three sphere and implementing certain change of variables associated with S-duality, we demonstrate that the data for the dual theory can be conveniently packaged in terms of a collection of framed affine $B_N$ quivers. In particular, the partition function for the dual of the $Sp(1)$ theory with $\kappa=1/2$ is
\begin{equation}
\CZ_{\text{dual}}= \CZ \left [\widehat{B}_N \right]+ {\CZ}\left [\widehat{B}'_N\right],
\label{eq:CZdualCS12}
\end{equation}
where $[\widehat{B}_N]$ and $[\widehat{B}'_N]$ are both framed affine $B_N$ quivers which differ by  the charge of the double arrow under the gauge groups $U(2)_{N-1} \times U(1)_{N}$ and the Chern-Simons level of the gauge group $U(1)_N$. 
%
Explicitly, one has
\begin{align}
\CZ\left[\widehat{B}_N\right]&=e^{-i\pi/4} \int d\mu\, \mathcal{Z}^{D}_{N-1}\, F^{(1)}_{\text{nsl}}\left(u_{N},u_{N-1}\right)\, \CZ^{\text{vec}}_{\text{bdry}} (u_{N},0,0), \cr
\CZ\left[\widehat{B}'_N\right]&=-e^{-i\pi/4} \int d\mu\, \mathcal{Z}^{D}_{N-1}\, F^{(2)}_{\text{nsl}}\left(u_{N},u_{N-1}\right)\, \CZ^{\text{vec}}_{\text{bdry}} \left(u_{N},-\frac{i}{2},-1\right)\,,
\label{eq:DualPfSummary}
\end{align}
where the subscript ``nsl'' indicates the contribution from the non-simply-laced edge of the quiver and the subscript ``bdry'' indicates the contribution from the boundary node associated with the short simple root of the $B_N$ algebra.  The other notations in the formulae are as follows: 
\ben
\item $d\mu$ is the appropriate measure of integration over the gauge group.
\item $\mathcal{Z}^{D}_{N-1}$ is the contribution from the $D_{N-1}$ quiver tail of a framed affine $B_N$ quiver whose explicit formula are given in \eref{eq:ZDN}.
\item $F^{(1,2)}_{\text{nsl}}(u_N,u_{N-1})$ depending on the Coulomb branch parameters of the last two nodes of the quiver $u_N,\, u_{N-1}$ are contributions of the double arrows for the framed affine ${B}_N$ and $B'_N$ quivers respectively.  The explicit formulae for these are given in \eref{BN1} and \eref{BN2}.
\item $\CZ^{\text{vec}}_{\text{bdry}}(u_N,\eta_N,\kappa)$ (see \eqref{eq:Zdefns} for the exact formula) is the contribution of the vector multiplet associated with the node of label $N$, which depends on the Coulomb branch parameter, the Fayet-Iliopoulos parameter and the Chern-Simons level (\figref{fig:quiverBN}). \Secref{sec:BNcomputation} contains details of this computation and related discussion.
\een

For a generic level $\kappa$ the dual theory partition function \eqref{eq:CZdualCS12} has $2\kappa+1$ terms with different Chern-Simons levels on the boundary node (see \eqref{eq:GenericCSlevel}).

\item \textbf{Contribution of  the double arrow} \\
The partition function computation allows us to read off the contributions of the double arrow connecting the $(N-1)$st and $N$th nodes of the framed affine $B_N$ quiver in $\CZ[\widehat{B}_N]$ and $\CZ[\widehat{B}'_N]$ respectively. For $k=1$ they are
\begin{align}
&F^{(1)}_{\text{nsl}}(u_N,u^l_{N-1}) = \frac{1}{\ch{(u_N-2 u^1_{N-1})}\ch{(u_N-2 u^2_{N-1})}}\,, \label{BN1} \\
& F^{(2)}_{\text{nsl}}(u_N,u^l_{N-1})= \frac{1}{\ch{(u_N-u^1_{N-1}+u^2_{N-1})}\ch{(u_N- u^2_{N-1}+u^1_{N-1})}}\label{BN2}\,.
\end{align}
Recall that an ordinary  hypermultiplet in a 3d $\CN=4$ theory contributes a factor of $\prod_{\rho} \frac{1}{\ch{\rho(u)}}$ to the integrand of an $S^3$ partition function, where the product is over all weights of the representation of the gauge group under which the given hypermultiplet transforms. The contribution of the double arrow in \eqref{BN1} has a similar form and one can therefore associate an ``effective weight" to the double arrow, i.e.
\begin{equation}
\rho^{\text{eff}}_{\text{da}} (u_N, u^i_{N-1})=u_N - 2u^i_{N-1}, \quad\quad i=1,2.
\end{equation}

Certainly the effective weight does not correspond to the weight of any representation of the gauge group. Note the factor of $2$ in the argument of $\cosh$ in \eqref{BN1} without which $F^{(1)}_{\text{nsl}}(u_N,u^i_{N-1})$ would be indistinguishable from the contribution of an ordinary bifundamental hyper. \\
The above formulae also show that the double arrow in $[\widehat{B}_N]$ is charged under the gauge group $U(2)_{N-1} \times U(1)_{N}$ while the double arrow in $[\widehat{B}'_N]$ is only charged under  $SU(2)_{N-1} \times U(1)_{N}$ where $SU(2)_{N-1}$ is a subgroup of $U(2)_{N-1}$.

\item \textbf{$\CN=4$ Superconformal Index ofthe framed affine $B_N$ quiver and its Coulomb branch limit} \\Using the effective weight associated with the double arrow, one can immediately conjecture a formula for the $\CN=4$ superconformal index of a framed affine $B_N$ quiver on $S^2 \times S^1$, since the contribution of matter multiplets to the index is also written as a product of weights. We discuss formula \eqref{eq:ProposedIndex} and its Coulomb branch limit in \Secref{sec:SupConfIndex4d}. We show that the Coulomb branch limit of the proposed index matches exactly with the Hilbert series of the moduli space of a single $B_N$ instanton on $\BC^2$. 
\end{itemize}

\subsection{Future Directions}
The analysis of the current work should be extended to the remaining non-simply laced quivers of CFG types which includes proper understanding of the construction from string theory as well as gauging the discrete symmetries of the corresponding ADE-type Dynkin diagrams.
The dictionary of the new duality (see \eqref{eq:MirrorMap0}), which generalizes mirror symmetry for gauge theories with nontrivial Chern-Simons terms, should be established in full generality for all non-simply laced series.

\section{The framed affine ${B}_{N}$ quivers from S-duality}\label{sec:BNcomputation}
In this section, we compute the partition function of an affine $B_{N}$ quiver with a single framing, as shown in \figref{fig:quiverBN}. Since there is no known Lagrangian description of such a theory, we cannot write down its partition function directly. Our strategy will be to start from the partition function of the mirror dual theory -- the ADHM quiver with $SO(2N+1)$ flavor symmetry and in the presence of half-integer Chern-Simons level $\kappa$. Then we shall perform S-duality and manipulate the resulting formula to obtain the partition function of the framed affine $B_{N}$ quiver. Since the computations are rather tedious we shall present the results for the relatively simpler case of one $SO(2N+1)$ instanton ($Sp(1)$ gauge theory).  Henceforth, we focus only on the case of $k=1$ in \figref{fig:quiverBN}.


\subsection{S-dualizing the Partition Function}
The partition function of $Sp(1)$ gauge theory at Chern-Simons level $\kappa$ with an $SO(2N+1)$ flavor symmetry and one antisymmetric hyper is \footnote{The $S^3$ partition function with a non-zero Chern-Simons term is not convergent. One needs to regularize the integral by adding a small positive imaginary piece to the Chern-Simons level and setting it to zero at the end of the computation. In the rest of the paper, we implicitly assume such a regularization.}
\begin{equation}
\begin{split}
\CZ_A&=\int \frac{d s}{2} \frac{\sinh^2{(2\pi s)} \cdot e^{2 \pi i \kappa s^2}}{\prod^N_{a=1}\cosh{\pi(s+m_a)}\cosh{\pi(s-m_a)}\cosh{\pi s}} \times \left(\frac{1}{\cosh{\pi M_{as}}}\right) \,,\label{pfBtriv-M=1}
\end{split}
\end{equation}
where the Cartan parameters of $Sp(1)$ are labelled by $\diag(s,-s)$, with a real number $s$ and the parameters for $SO(2N+1)$ are taken to be $\diag(m_a,-m_a,0)$, with real numbers $m_a, a=1\ldots N$. Hypermultiplets transform in the bi-fundamental representation of $Sp(1) \times SO(2N+1)$ as one can clearly see from the structure of the integrand. The antisymmetric hypermultiplet for $Sp(1)$ of mass $M_{as}$ is a singlet and the contribution of this singlet in the partition function is given by the last term in parenthesis, indicating that it can be factored out of the integration.

The computation is rather technical and tedious, we therefore describe it in full detail in Appendices \ref{Sec:CauchyFourier}, \ref{YZ} and \ref{Sec:Znsl}. Here let us merely outline the strategy and write down the results. First we apply the Cauchy determinant identity to the integrand of \eqref{pfBtriv-M=1}, which will reshape the expression to be better suitable for the Fourier transform. The latter, similarly to the known examples of mirror dual quiver theories of A-type \cite{Kapustin:2010xq}, manifest the duality transformation. Then, after an appropriate change of variables, the integral can be regarded as a partition function of the dual theory with $B_N$ symmetry.

The resulting expression for $\kappa=\frac{1}{2}$ \footnote{A schematic form of the formula for a generic half-integer $\kappa$ is given in \eqref{eq:GenericCSlevel}.} reads
\bea
\CZ_{B}= \CZ [\widehat{B}_N]+ {\CZ}[\widehat{B}'_N]\,,
\label{eq:ZBtwosectors}
\eea
which depends on FI parameters $\eta_0\,,\dots,\eta_N$ of the gauge nodes of the framed affine $B_N$ quiver. Below we specify this dependences in full detail. The constituents of the right hand side of \eqref{eq:ZBtwosectors} are given by the following integrals
\begin{align}
\CZ[\widehat{B}_N]&={e^{-i\pi/4}} \int d\mu\, \mathcal{Z}^{D}_{N-1}\, F^{(1)}_{\text{nsl}}(u_{N},u^p_{N-1})\, \CZ^{\text{vec}}_{\text{bdry}} (u_{N},0,0), \cr
\CZ[\widehat{B}'_N]&=-e^{-i\pi/4} \int d\mu\, \mathcal{Z}^{D}_{N-1}\, F^{(2)}_{\text{nsl}}(u_{N},u^p_{N-1})\, \CZ^{\text{vec}}_{\text{bdry}} \left(u_{N},-\frac{i}{2},-1\right)\,,
\label{ZBN} 
\end{align}
in which the measure of integration is
\begin{equation}
d\mu = \frac{1}{(2!)^{N-2}} \prod^{1}_{\alpha=0}d u_{\alpha} d u_{N} \prod^{N-1}_{\beta=2}d^{2} {u}_{\beta}\,.
\end{equation}
The contribution of vector multiplets for nodes 1 through $N-2$ and hypermultiplets connecting them (the $D$-shaped left side of the quiver in \figref{fig:quiverBN}) reads as
\begin{equation}
\mathcal{Z}^{D}_{N-1} = \frac{\CZ^{\text{vec}}_{\text{bdry}} (u_0,\eta_0,0)}{\CZ^{\text{bif}}_{\text{bdry}} (u_0, u_{2}) \CZ^{\text{fund}}_{\text{bdry}} (u_0)} \times  \frac{\CZ^{\text{vec}}_{\text{bdry}} (u_1,\eta_1,0)}{\CZ^{\text{bif}}_{\text{bdry}} (u_1, u_{2})}\times \frac{\prod^{N-1}_{\beta=2}\CZ^{\text{vec}} (u_{\beta},{\eta}_\beta,0) }{\prod^{N-2}_{\beta=2} \CZ^{\text{bif}}(u_{\beta},u_{\beta+1})}\,, 
\label{eq:ZDN}
\end{equation}
and the novel contributions for the matter corresponding to the double arrow $F^{(1,2)}_{\text{nsl}}$ and the vector multiplet on the right-most node of the quiver $\CZ^{\text{vec}}_{\text{bdry}}$ are given below
\begin{equation}
\begin{split}
&F^{(1)}_{\text{nsl}}(u_N,u^l_{N-1}) = \frac{1}{\ch{(u_N-2 u^1_{N-1})}\ch{(u_N-2 u^2_{N-1})}}\,,\\
& F^{(2)}_{\text{nsl}}(u_N,u^l_{N-1})= \frac{1}{\ch{(u_N-u^1_{N-1}+u^2_{N-1})}\ch{(u_N- u^2_{N-1}+u^1_{N-1})}}\,,
\label{eq:Fnsldef}
\end{split}
\end{equation}
where the subscript ``nsl'' indicates the contribution from the non-simply-laced edge of the quiver and the subscript ``bdry'' indicates the contribution from the boundary node associated with the short simple root of the $B_N$ algebra.

The various perturbative contributions of \eqref{ZBN} and \eqref{eq:ZDN}  are
\begin{equation}
\begin{split}
&\CZ^{\text{vec}}_{\text{bdry}}(u,\eta,\kappa)= e^{2\pi i \eta u}e^{\pi i \kappa (u)^2},\\
&\CZ^{\text{bif}}_{\text{bdry}}(u,\vec{v})=\prod^2_{p=1} \ch{(u-v^p)},\\
&\CZ^{\text{fund}}_{\text{bdry}}(u)=\ch{u},\\
&\CZ^{\text{vec}} (\vec{u},\eta, \kappa)=\sinh^2{\pi({u}^1-{u}^2)} \prod^{2}_{p=1}e^{2\pi i \eta {u}^p} e^{\pi i \kappa ({u}^p)^2},\\
&\CZ^{\text{bif}}(\vec{u},\vec{v})=\prod^2_{p,l=1}\cosh{\pi({u}^p -{v}^l)}.
\label{eq:Zdefns}
\end{split}
\end{equation}
The dual partition function $\CZ_{B}$ can therefore be written as a sum of two contributions each representing a partition function for
a $\widehat{B}_N$-type quiver theory (having gauge group $U(2)^{N-2} \times U(1)^3$ and appropriate matter fields), where the contributions of the double arrow are given by functions $F^{(1)}_{\text{nsl}}(u_N,u^l_{N-1})$ and $F^{(2)}_{\text{nsl}}(u_N,u^l_{N-1})$ respectively.

Note that in $\CZ [\widehat{B}_N]$ the matter corresponding to the double arrow is charged under $U(1)_N \times U(2)_{N-1}$, while in ${\CZ}[\widehat{B}'_N]$ this matter is charged under $U(1)_N \times SU(2)_{N-1}$ but not under the $U(1)$ subgroup of $U(2)_{N-1}$. 

In other words, partition function $\CZ[\widehat{B}_N]$ can be obtained by a simple deformation of the partition function $\CZ[\widehat{D}_{N+1}]$ of the framed affine ${D}_{N+1}$ quiver:
\begin{equation}
\CZ[\widehat{D}_{N+1}]=\int  d\mu\, \mathcal{Z}^{D}_{N-1} \cdot \frac{\CZ^{\text{vec}}_{\text{bdry}} (u_{N},\eta_N) }{\CZ^{\text{bif}}_{\text{bdry}}(u_{N},u_{N-1})}.
\end{equation}
by the following deformation ($\mathbb{Z}_2$ folding)
\begin{equation}
\begin{split}
\CZ^{\text{bif}}_{\text{bdry}} (u_N, u_{N-1})= \frac{1}{\prod^2_{p=1} \ch{(u_{N}-u^p_{N-1})}}\to 
\frac{1}{\mathcal{Z}_{\text{nsl}}(u_{N},u^p_{N-1};\kappa)}.
\end{split}
\end{equation}
where $\mathcal{Z}_{\text{nsl}}$ is given by \eref{eq:BoxedFnsl}.
As is evident, $\mathcal{Z}_{\text{nsl}}$ cannot be obtained as a product over weights of any representation of gauge group $U(2)_{N-1} \times U(1)_N$.

\subsection{The Duality Map} 
Three-dimensional mirror symmetry interchanges Fayet-Iliopoulos parameters and masses of the two dual theories. Expectedly this happens for our duality as well, so the first part of the dictionary reads 
\begin{equation}
\label{eq:DictionarymassesfI}
\begin{split}
&\eta_0=-M_{as} -(m_{1}+m_2)\,, \\
&\eta_\beta=m_{\beta} -m_{\beta +1 }\,,\qquad  \beta=1,\ldots,N-2\,,\\
&\eta_{N-1}=m_{N-1},\\
&\eta_N=0\,.
\end{split}
\end{equation}

If we neglect the Chern-Simons terms and set $\kappa=0$, then the second term in \eqref{eq:ZBtwosectors} vanishes and \eqref{eq:DictionarymassesfI} describes the complete map of the parameters. However, due to the presence of the Chern-Simons term in the $Sp(1)$ theory the above dictionary needs to be completed by some extra data -- the framed affine $B_3$ theory also has its Chern-Simons couplings according to \eqref{ZBN}. In particular, $\CZ[\widehat{B}'_N]$ contains Chern-Simons  level $\kappa_N=-1$.

\subsection{Dual Partition Function for generic Chern-Simons Levels}
So far we have only studied the duality for $\kappa=1/2$, however, we have already derived the expression for generic level in \eqref{nsl0}.
In order to interpret the result in terms of the framed affine $B_N$ quiver, we expand the relevant part of the integrand as
\begin{equation}
\frac{\sinh{2\pi\kappa s}}{\sh{s}} = e^{-(2\kappa-1)\pi s}+e^{-(2\kappa-2)\pi s}+\dots+e^{(2\kappa-1)\pi s}\,,
\end{equation}
we generate $2\kappa+1$ terms for the dual partition function
\begin{equation}
\CZ_B = \CZ[\widehat{B}_N]+\sum_{i=1}^{2\kappa}\CZ[\widehat{B}^{(i)}_N]\,,
\label{eq:GenericCSlevel}
\end{equation}
with the same (up to a prefactor) term $\CZ[\widehat{B}_N]$ as in \eqref{eq:ZBtwosectors} and with $2\kappa$ terms which with different Chern-Simons levels on the $N$-th node. These terms vanish as we put the Chern-Simons coupling to zero.

\section{$\CN=2$ superconformal index in 4d and the $\CN=4$ $S^2 \times S^1$ index}\label{sec:SupConfIndex4d}
In the previous section we have presented an explicit expression for the supersymmetric partition function \eqref{eq:ZBtwosectors} of the non-Lagrangian theory which is given by the framed affine $B_N$ quiver \figref{fig:quiverBN}. We have derived the desired result by studying S-dual configuration of the ADHM quiver for the moduli space of a single $SO(2N+1)$ instanton. In this section we provide further evidence which supports our proposal.

Let us first consider the simpler case of the anomalous theory ($\kappa=0$) with $N=3$ with $k=1$. This theory and its mirror dual have $\CN=4$ supersymmetry -- therefore one should compute the $\CN=4$ $S^2 \times S^1$ index in this case. In this section, we give a prescription for formally writing down the $\CN=4$ $S^2 \times S^1$ index of the framed affine $B_{3}$ quiver. Although this is the index of an unphysical theory (mirror of an anomalous theory), one can use it to demonstrate that the Coulomb branch limit of the index agrees exactly with the Hilbert series of a $B_3$ instanton on $\mathbb{C}^2$ -- this reproduces the result found in \cite{Cremonesi:2014xha}. However, the Higgs branch limit of the index of the framed affine $B_{3}$ quiver cannot be interpreted as a generating function of operators on some moduli space.
\begin{figure}[ht]
\begin{center}
\includegraphics[scale=0.4]{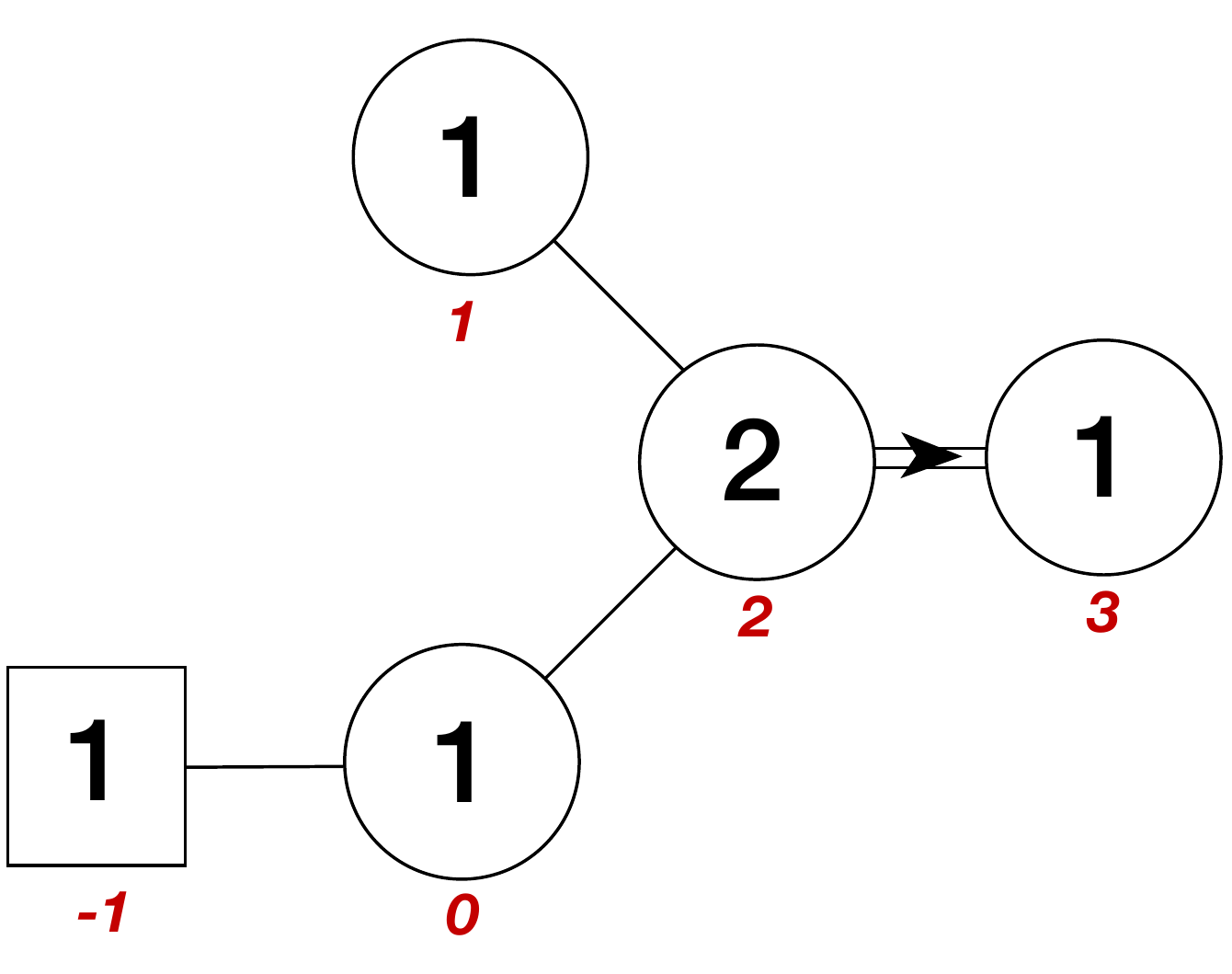}
\caption{The framed affine $B_{3}$ quiver, also denoted by $[\widehat{B}_3]$, with labels.}
\label{fig:quiverbN}
\end{center}
\end{figure}
The 3d partition function is of the following form:
\begin{align}
\CZ[\widehat{B}_3]=\int  \prod^{1}_{\alpha=0}d s^{(\alpha)} d s^{(3)} \frac{d^{2} {s}^{(2)}}{2!}  \times \frac{\CZ^{\text{vec}}_{\text{bdry}} (s^{(0)})}{\CZ^{\text{bif}}_{\text{bdry}} (s^{(0)}, s^{(2)}) \CZ^{\text{fund}}_{\text{bdry}} (s^{(0)})} \times  \frac{\CZ^{\text{vec}}_{\text{bdry}} (s^{(1)})}{\CZ^{\text{bif}}_{\text{bdry}} (s^{(1)}, s^{(2)})} \\ \nonumber
\times \CZ^{\text{vec}} (s^{(2)}) \times \CZ^{\text{nsl}}(s^{(2)},s^{(3)})  \times \CZ^{\text{vec}}_{\text{bdry}} (s^{(3)}).
\label{eq:indexdef}
\end{align}
Note that the contribution of the matter part of the partition function can still be written as $\prod_{\rho} \frac{1}{\ch{\rho(s)}}$ -- where the product goes over all weights of the representation of the gauge group under which a given matter multiplet transforms -- provided we associate an ``effective weight" with the double arrow, i.e.
\begin{equation}
\rho_{\text{da}}(s^{(2)},s^{(3)})=s^{(3)}- 2s^{(2)}_i, \quad i=1,2.
\end{equation}
This immediately suggests how the formula for the 4d index (where contribution of matter multiplets is also written as a product of weights as above) should be modified  for the framed affine $B_3$ quiver: we treat the double bond as a multiplet with these `effective weights' in the index formula.

\subsection{Lens Space index of the framed affine $B_{3}$ quiver}
Most terms in the 3d partition function of the framed affine $B_{3}$ quiver can be readily identified as the contributions of vector and hyper multiplets -- the only exception being the contribution of the double arrow in the quiver. Writing the Lens space index of the theory simply involves replacing the vector and hyper contributions by the appropriate indices (given above) and replacing the function $\mathcal{Z}_{\text{nsl}}(s^{(2)},s^{(3)})$ by a deformed function $\mathcal{I}_{\text{nsl}}(z^{(2)},z^{(3)};r)$. A summary of superconformal indices on Lens spaces as partition functions on $S^3\times S^1$ is presented in \appref{Sec:GenIndex}. 

In the limit when $S^1$ shrinks the contribution of the double arrow to the superconformal index should reduce to the corresponding term in the $S^3$ partition function which we have studied above 
\begin{equation}
\mathcal{I}_{\text{nsl}}(z^{(2)},z^{(3)};r) \to \mathcal{Z}_{\text{nsl}}(s^{(2)},s^{(3)})\,,
\end{equation}
where $z^{(2)}_i =e^{2\pi i s^{(2)}_i }$. Therefore the full index should have the following form
\begin{align}
\mathcal{I}(p,q,t;\widetilde{z}^{(\alpha)}, \widetilde{m}^{(\alpha)})=\sum_{\{m^{(\alpha)}\}} \oint\limits_{|z^{(\alpha)}_i|=1} \frac{dz^{(0)}}{2\pi i z^{(0)}} \frac{dz^{(1)}}{2\pi i z^{(1)}} \frac{dz^{(3)}}{2\pi i z^{(3)}} \frac{1}{W(m^{(2)})} \prod_{i=1,2} \frac{dz^{(2)}_i}{2\pi i z^{(2)}_i}  \\ \nonumber
\times f(p,q,t;r) \times \mathcal{I}_{\text{fund}}^{(m^{(0)}, \widetilde{m}^{(-1)})} (z^{(0)}, \widetilde{z}^{(0)}) \mathcal{I}_\text{bifund}^{(m^{(0)}, m^{(2)})} (z^{(0)}, z^{(2)})\mathcal{I}_{\text{bifund}}^{(m^{(1)}, m^{(2)})} (z^{(1)}, z^{(2)}) \\ \nonumber
\times \mathcal{I}_V^{(m^{(0)})}(z^{(0)}) \mathcal{I}_V^{(m^{(1)})}(z^{(1)}) \mathcal{I}_V^{(m^{(2)})}(z^{(2)}) \times  \mathcal{I}^{(m^{(3)}, m^{(2)})} _{\text{nsl}}(z^{(3)},z^{(2)};r) \mathcal{I}_V^{(m^{(3)})}(z^{(3)})\,,
\end{align}
where $W(m^{(2)})$ is the order of the Weyl group of the gauge group preserved by a given $\{m^{(2)}_i\}$ -- i.e. $W(m^{(2)})=2!$ if $m^{(2)}_1 \neq m^{(2)}_2$ and $W(m^{(2)})=1$ if $m^{(2)}_1 = m^{(2)}_2$. At the moment we can define the index above up to an arbitrary function $f(p,q,t;r)$ of flavor fugacities, which we shall be able to fix later in this section.

The individual functions appearing in the index above are given as:
\begin{equation}
\begin{split}
&\mathcal{I}_\text{fund}^{(m^{(0)}, \widetilde{m}^{(-1)})} (z^{(0)}, \widetilde{z}^{(0)})=\left(\frac{pq}{t}\right)^{\frac{1}{2}([[(m^{(0)}- \widetilde{m}^{(-1)})]]-\frac{1}{r}[[(m^{(0)}, \widetilde{m}^{(-1)})]]^2 )}\\
& \times \prod_{s=\pm 1} \Gamma(t^{1/2} p^{[[s(m^{(0)}-\widetilde{m}^{(-1)})]]} e^{2\pi i s (s^{(0)} -\widetilde{s}^{(0)})};pq,p^r) \Gamma(t^{1/2} q^{r-[[s(m^{(0)} -\widetilde{m}^{(-1)})]]}e^{2\pi i s (s^{(0)} -\widetilde{s}^{(0)})} ;pq,q^r),\\
&\mathcal{I}_\text{bifund}^{(m^{(\alpha)}, m^{(2)})} (z^{(\alpha)}, z^{(2)})=\prod_{i}\left(\frac{pq}{t}\right)^{\frac{1}{2}([[(m^{(\alpha)} - m_i^{(2)})]]-\frac{1}{r}[[(m^{(\alpha)} - m_i^{(2)})]]^2 )}\\
& \times \prod_{s=\pm 1} \Gamma(t^{1/2} p^{[[s(m^{(\alpha)} - m_i^{(2)})]]} e^{2\pi i s (s^{(\alpha)}-s_i^{(2)})};pq,p^r) \Gamma(t^{1/2} q^{r-[[s(m^{(\alpha)}-m_i^{(2)})]]}e^{2\pi i s (s^{(\alpha)}- s_i^{(2)})} ;pq,q^r),\\
&\mathcal{I}_V^{(m^{(i)})}(z^{(i)})=\frac{(p^r;p^r)}{\Gamma(t;pq,p^r)} \frac{q^r;q^r}{\Gamma(tq^r;pq,q^r)} \;\;(i=0,1,3),\\
& \mathcal{I}_V^{(m^{(2)})}(z^{(2)})=\left(\frac{(p^r;p^r)}{\Gamma(t;pq,p^r)} \frac{q^r;q^r}{\Gamma(tq^r;pq,q^r)}\right)^{2} \prod_{i\neq j}\left(\frac{pq}{t}\right)^{-\frac{1}{2}([[(m_i^{(2)} -m_j^{(2)})]]-\frac{1}{r}[[(m_i^{(2)} -m_j^{(2)})]]^2 )}\\
& \times \frac{1}{\Gamma(tp^{[[(m_i^{(2)} -m_j^{(2)})]]} e^{2\pi i (s^{(2)}_i -s^{(2)}_j)};pq,p^r)} \frac{1}{\Gamma(tq^{r-[[(m_i^{(2)} -m_j^{(2)})]]} e^{2\pi i (s_i^{(2)} - s_j^{(2)})};pq,p^r)}\\
& \times \frac{1}{\Gamma(p^{[[(m_i^{(2)} -m_j^{(2)})]]} e^{2\pi i (s_i^{(2)} - s_j^{(2)})};pq,p^r)} \frac{1}{\Gamma(q^{r-[[(m_i^{(2)} -m_j^{(2)})]]} e^{2\pi i (s^{(2)}_i -s^{(2)}_j)};pq,p^r)},\\
&\mathcal{I}^{(m^{(3)}, m^{(2)})} _{\text{nsl}}(z^{(3)},z^{(2)};r)=\prod_{i}\left(\frac{pq}{t}\right)^{\frac{1}{2}([[(m^{(3)} -2 m_i^{(2)})]]-\frac{1}{r}[[(m^{(3)} - 2m_i^{(2)})]]^2 )}\\
& \times \prod_{s=\pm 1} \Gamma(t^{1/2} p^{[[s(m^{(3)} - 2m_i^{(2)})]]} e^{2\pi i s (s^{(3)}-2 s_i^{(2)})};pq,p^r) \Gamma(t^{1/2} q^{r-[[s(m^{(3)}-2 m_i^{(2)})]]}e^{2\pi i s (s^{(3)}- 2 s_i^{(2)})} ;pq,q^r).
\end{split}
\end{equation}

Note that the last line is the proposed form of the contribution of the double arrow in the framed affine $B_3$ quiver to the Lens space index. For a generic case the prescription is simply:
\begin{equation}
\begin{split}
&\mathcal{I}_{\text{nsl}}^{(m^{(\beta)}, m^{(\gamma)})} (z^{(\beta)}, z^{(\gamma)})=\prod_{\rho}\left(\frac{pq}{t}\right)^{\frac{1}{2}([[\rho(m^{(\beta)}, m^{(\gamma)})]]-\frac{1}{r}[[\rho(m^{(\beta)}, m^{(\gamma)})]]^2 )}\\
& \times \prod_{s=\pm 1} \Gamma(t^{1/2} p^{[[s\rho(m^{(\beta)}, m^{(\gamma)})]]} e^{2\pi i s \rho(s^{(\beta)},s^{(\gamma)})};pq,p^r) \Gamma(t^{1/2} q^{r-[[s\rho(m^{(\beta)}, m^{(\gamma)})]]}e^{2\pi i s \rho(s^{(\beta)},s^{(\gamma)})} ;pq,q^r);\\
&\rho(m^{(\beta)}, m^{(\gamma)}) =\{m^{(\beta)}_i - 2m^{(\gamma)}_j | \forall i,j\},\;\rho(s^{(\beta)}, s^{(\gamma)}) =\{s^{(\beta)}_i - 2s^{(\gamma)}_j | \forall i,j\},
\end{split}
\end{equation}
for a double bond between $U(N_\beta)$ and $U(N_\gamma)$ with the arrow directed from the node $(\gamma)$ to node $(\beta)$.

\subsection{Projection to $S^2 \times S^1$ index}
Consider the following redefinition of fugacities in the Lens space index:
\begin{equation}
\begin{split}
p=\tq ^{1/2} y,\quad q=\tq^{1/2} y^{-1}, \quad t=\tti \tq^{1/2}
\end{split}
\end{equation}
Under the above redefinition, the index in \eqref{4dindex-def1} can be written as
\begin{equation}\label{4dindex-def2}
\begin{split}
\mathcal{I}(\tq,y,\tti;z_i)= \tr_{S^3/\mathbb{Z}_r} \left[(-1)^F (\tq)^{j_2 + \frac{R-R'}{2}} (\tti)^{R+R'} y^{2j_1} e^{-\beta(E-2j_2 -2R +R')} \prod_i z_i^{f_i}\right].
\end{split}
\end{equation}
The $S^2\times S^1$ index can now be defined as $r\to \infty$ limit of the lens index (see \cite{Razamat:2014aa})
\begin{equation}
\begin{split}
\mathcal{I}_{S^2\times S^1}=& \lim_{r\to \infty} \mathcal{I}(\tq,y,\tti;z_i)\vert_{y=1}\\
=& \tr_{S^2} \left[(-1)^F (\tq)^{j_2 + \frac{R-R'}{2}} (\tti)^{R+R'}  e^{-\beta(E-2j_2 -2R +R')} \prod_i z_i^{f_i}\right].
\end{split}
\end{equation}
Now, since the index has non-zero contributions from only those states which satisfy $E-2j_2-2R+R'=0$, one may rewrite the 3d conformal dimension $\tE=\frac{E-R'}{2}$ for these states as
\begin{equation}
\tE=j_2 +R-R'.
\end{equation}
In terms of $\tE$, the 3d index can be written as 
\begin{equation}
\begin{split}
&\mathcal{I}_{S^2\times S^1} (\tq,\tti;z_i)= \tr_{S^2} \left[(-1)^F (x)^{\tE +R'} (\tx)^{\tE -R} e^{-\beta(\tE-j_2 -R +R')} \prod_i z_i^{f_i} \right],\\
&x=\tq^{1/2} \tti, \quad \tx=\tq^{1/2} \tti^{-1}.
\end{split}
\end{equation}
There are two useful limits of the 3d index that we will often use -- the Coulomb branch index $\mathcal{I}_C$ and the Higgs branch index $\mathcal{I}_H$ which are defined as follows:
\begin{equation}
\begin{split}
&\mathcal{I}_C=\tr_{\CH_C} \left[(-1)^F(\tx)^{\tE -R} e^{-\beta(\tE-j_2 -R +R')} \prod_i z_i^{f_i} \right]=\lim_{x \to 0} \mathcal{I}_{S^2\times S^1}(x,\tx; z_i)\\
&\mathcal{I}_H=\tr_{\CH_H} \left[(-1)^F(x)^{\tE +R'} e^{-\beta(\tE-j_2 -R +R')} \prod_i z_i^{f_i} \right]=\lim_{\tx \to 0} \mathcal{I}_{S^2\times S^1}(x,\tx; z_i)
\end{split}
\end{equation}
where $\CH_C$ is the subspace of the Hilbert space where states satisfy $\tE +R'=0$ and $\CH_H$ is the subspace of the Hilbert space where states satisfy $\tE -R=0$.\\

Now let us write down the proposed 3d index for the framed affine $B_3$ quiver.
\begin{equation}
\begin{split}
&\mathcal{I}(\tq,\tti;\widetilde{z}^{(\alpha)}, \widetilde{m}^{(\alpha)})=g(\tq,\tti)\sum_{\{m^{(\alpha)}\}} \oint_{|z^{(\alpha)}_i|=1} \frac{dz^{(0)}}{2\pi i z^{(0)}} \frac{dz^{(1)}}{2\pi i z^{(1)}} \frac{dz^{(3)}}{2\pi i z^{(3)}} \frac{1}{W(m^{(2)})} \prod_{i=1,2} \frac{dz^{(2)}_i}{2\pi i z^{(2)}_i} \\
& \times\mathcal{I}_V^{(m^{(0)})}(z^{(0)}) \mathcal{I}_V^{(m^{(1)})}(z^{(1)})\mathcal{I}_\text{fund}^{(m^{(0)}, \widetilde{m}^{(-1)})} (z^{(0)}, \widetilde{z}^{(0)}) \mathcal{I}_\text{bifund}^{(m^{(0)}, m^{(2)})} (z^{(0)}, z^{(2)})\mathcal{I}_\text{bifund}^{(m^{(1)}, m^{(2)})} (z^{(1)}, z^{(2)}) \\
&\times \mathcal{I}_V^{(m^{(2)})}(z^{(2)}) \mathcal{I}^{(m^{(3)}, m^{(2)})} _{\text{nsl}}(z^{(3)},z^{(2)}) \mathcal{I}_V^{(m^{(3)})}(z^{(3)}),
\end{split}
\label{eq:ProposedIndex}
\end{equation}
where $g(\tq,\tti) = \lim\limits_{r \to \infty} f(p,q,t;r)$ and the other ingredients of the above equation are:
\begin{align} 
&\mathcal{I}_\text{fund}^{(m^{(0)}, \widetilde{m}^{(-1)})} (z^{(0)}, \widetilde{z}^{(0)})=(\frac{\tq^{1/2}}{\tti})^{\frac{1}{2} |m^{(0)}-\widetilde{m}^{(-1)}|} \frac{(\tti^{-1/2} \tq^{3/4+|m^{(0)}-\widetilde{m}^{(-1)}|/2} (z^{(0)}/\widetilde{z}^{(0)})^{\pm 1};\tq)}{(\tti^{1/2} \tq^{1/4+|m^{(0)}-\widetilde{m}^{(-1)}|/2} (z^{(0)}/\widetilde{z}^{(0)})^{\pm 1};\tq)}, \label{sci_fund}\\
& \mathcal{I}_\text{bifund}^{(m^{(\alpha)}, {m}^{(2)})} (z^{(\alpha)}, {z}^{(2)})=\prod_{i=1,2}(\frac{\tq^{1/2}}{\tti})^{\frac{1}{2} |m^{(\alpha)}-{m}_i^{(2)}|} \frac{(\tti^{-1/2} \tq^{3/4+|m^{(\alpha)}-{m}_i^{(2)}|/2} (z^{(\alpha)}/{z}_i^{(2)})^{\pm 1};\tq)}{(\tti^{1/2} \tq^{1/4+|m^{(\alpha)}-{m}_i^{(2)}|/2} (z^{(\alpha)}/{z}_i^{(2)})^{\pm 1};\tq)},\label{sci_bifund}\\
&\mathcal{I}_V^{(m^{(i)})}(z^{(i)}) =\frac{(\tti \tq^{1/2};\tq)}{(\tti^{-1} \tq^{1/2};\tq)} \;\; (i=0,1,3)\label{sci_abV},\\
&\mathcal{I}_V^{(m^{(2)})}(z^{(2)}) =\left(\frac{(\tti \tq^{1/2};\tq)}{(\tti^{-1} \tq^{1/2};\tq)}\right)^2 \prod_{i \neq j} (\frac{\tq^{1/2}}{\tti})^{-\frac{1}{2} |m_i^{(2)}-{m}_j^{(2)}|}\frac{(\tti \tq^{1/2+|m_i^{(2)}-{m}_j^{(2)}|/2} z_i^{(2)}/{z}_j^{(2)};\tq) }{(\tti^{-1} \tq^{1/2+|m_i^{(2)}-{m}_j^{(2)}|/2} z_i^{(2)}/{z}_j^{(2)};\tq)}\nn \\
& \times(1-\tq^{\frac{1}{2}|m^{(2)}_i -m_j^{(2)}|}z_i^{(2)}/{z}_j^{(2)} ), \label{sci_nonabV}\\
&\mathcal{I}^{(m^{(3)}, m^{(2)})} _{\text{nsl}}(z^{(3)},z^{(2)})= \prod_{i=1,2}(\frac{\tq^{1/2}}{\tti})^{\frac{1}{2} |m^{(3)}-2{m}_i^{(2)}|} \frac{(\tti^{-1/2} \tq^{3/4+|m^{(3)}-2{m}_i^{(2)}|/2} (z^{(3)}/({z}_i^{(2)})^2)^{\pm 1};\tq)}{(\tti^{1/2} \tq^{1/4+|m^{(\alpha)}-2{m}_i^{(2)}|/2} (z^{(3)}/({z}_i^{(2)})^2)^{\pm 1};\tq)}. \label{sci_nsl1}
\end{align}


\subsection{Coulomb branch index of the framed affine $B_3$ theory}
In the limit $x \to 0$ and $\widetilde x$ is fixed, various factors in \eref{sci_fund}--\eref{sci_nsl1} reduce to the following forms:
\begin{equation}
\begin{split}
& \mathcal{I}_\text{fund}^{(m^{(0)}, \widetilde{m}^{(-1)})} (z^{(0)}, \widetilde{z}^{(0)}) \to \tx ^{\frac{1}{2} |m^{(0)}-\widetilde{m}^{(-1)}|}\\
& \mathcal{I}_\text{bifund}^{(m^{(\alpha)}, {m}^{(2)})} (z^{(\alpha)}, {z}^{(2)}) \to \prod_{i=1,2}(\tx)^{\frac{1}{2} |m^{(\alpha)}-{m}_i^{(2)}|}~ ,\quad \alpha =0,1\\
& \mathcal{I}_V^{(m^{(2)})}(z^{(2)}) \to \left\{
     \begin{array}{lr}
       (1-\tx)^{-2} \prod_{i \neq j}  (\tx)^{-\frac{1}{2}|m_i^{(2)}-{m}_j^{(2)}|} & : m^{(2)}_1 \neq m^{(2)}_2\\
        (1-\tx)^{-2}\prod_{i \neq j} (1-\frac{z_i^{(2)}}{z_j^{(2)}})/(1-\tx \frac{z_i^{(2)}}{z_j^{(2)}}) & : m^{(2)}_1= m^{(2)}_2
     \end{array}
   \right.\\
& \mathcal{I}^{(m^{(3)}, m^{(2)})} _{\text{nsl}}(z^{(3)},z^{(2)}) \to \prod_{i=1,2}(\tx)^{\frac{1}{2} |m^{(3)}-2{m}_i^{(2)}|} \\
& \mathcal{I}_V^{(m^{(i)})}(z^{(i)}) \rightarrow \frac{1}{1-\widetilde{x}}~, \quad  i =0,1,3~.
\end{split}
\end{equation}

Therefore, the Coulomb branch index can be written as
\begin{equation}
\begin{split} \label{CBI1}
&g^{-1}(\widetilde{x},x=0)\mathcal{I}_C(\tx; m^{(-1)}) = S_1 + S_2\\
&=\sum_{\{m^{(\alpha)}, m^{(2)}_1= m^{(2)}_2\}} \oint_{|z^{(\alpha)}_i|=1} \prod_{\alpha=0,1,3}\frac{dz^{(\alpha)}}{2\pi i z^{(\alpha)}} \prod_{i=1,2} \frac{dz^{(2)}_i}{2\pi i z^{(2)}_i} \tx ^{\frac{1}{2} |m^{(0)}-\widetilde{m}^{(-1)}|} \prod_{\alpha=0,1} \prod_{i=1,2}(\tx)^{\frac{1}{2} |m^{(\alpha)}-{m}_i^{(2)}|}\\
&\times (1-\tx)^{-3} \prod_{i \neq j} (1-\frac{z_i^{(2)}}{z_j^{(2)}})/(1-\tx \frac{z_i^{(2)}}{z_j^{(2)}}) \prod_{i=1,2}(\tx)^{\frac{1}{2} |m^{(3)}-2{m}_i^{(2)}|}\\
& + \sum_{\{m^{(\alpha)}, m^{(2)}_1\neq m^{(2)}_2\}} \oint_{|z^{(\alpha)}_i|=1} \prod_{\alpha=0,1,3}\frac{dz^{(\alpha)}}{2\pi i z^{(\alpha)}} (\frac{1}{2!})\prod_{i=1,2} \frac{dz^{(2)}_i}{2\pi i z^{(2)}_i} \tx ^{\frac{1}{2} |m^{(0)}-\widetilde{m}^{(-1)}|} \prod_{\alpha=0,1} \prod_{i=1,2}(\tx)^{\frac{1}{2} |m^{(\alpha)}-{m}_i^{(2)}|}\\
&\times (1-\tx)^{-3} \prod_{i \neq j}  (\tx)^{-\frac{1}{2}|m_i^{(2)}-{m}_j^{(2)}|}  \prod_{i=1,2}(\tx)^{\frac{1}{2} |m^{(3)}-2{m}_i^{(2)}|}.
\end{split}
\end{equation}
The RHS is in fact equal to to the Hilbert series of the moduli space of one $B_3$ instanton on $\BC^2$. We next show that this is indeed the case.

Define $\tx=t^2$, then the individual indices are
\begin{align} 
\CI_{\rm fund}^{(m^{(0)},\widetilde{m}^{(-1)})}  \quad &\rightarrow \quad t^{|m^{(0)} - \widetilde{m}^{(-1)}|}\\
\CI_{\rm bifund}^{(m^{(\alpha)},\widetilde{m}^{(-1)})}  \quad &\rightarrow \quad \prod_{i=1,2} t^{|m^{(\alpha)} - \widetilde{m}^{(2)}_i|}~, \quad \alpha=0,1 \\
\CI_{\rm V}^{(m^{(2)})}  \quad &\rightarrow \quad \begin{cases}  (1-t^2)^{-2} \;   t^{-2|m^{(2)}_1 -m^{(2)}_2|} &\qquad m^{(2)}_1 \neq m^{(2)}_2\\ \\  (1-t^2)^{-2} \; t^{-2|m^{(2)}_1 -m^{(2)}_2|} \prod_{i\neq j} \left( 1-\frac{z^{(2)}_i}{z^{(2)}_j} \right) /\left( 1- t^2 \frac{z^{(2)}_i}{z^{(2)}_j} \right)  \\  \qquad = (1-t^2)^{-2} \left( 1-\frac{z^{(2)}_i}{z^{(2)}_j} \right) /\left( 1- t^2 \frac{z^{(2)}_i}{z^{(2)}_j} \right) &\qquad m^{(2)}_1 = m^{(2)}_2\end{cases} \\
\CI_{\rm \text{nsl}}^{(m^{(3)},m^{(2)})} \quad &\rightarrow \quad \prod_{i=1,2} t^{|2m^{(2)}_i-m^{(3)}|} \\
 \mathcal{I}_V^{(m^{(i)})}(z^{(i)})\quad & \rightarrow  \quad \frac{1}{1- t^2}~, \quad  i=0,1,3~. \label{teqCoulomb}
\end{align}
The integrations over the gauge fugacities $z^{(0)}$, $z^{(1)}$, $z^{(3)}$ and $z^{(2)}$ when $m_1 \neq m_2$ are trivial while that over $z^{(2)}$ when $m_1 = m_2$ can be performed easily:
\begin{align}
&\frac{1}{2!}\frac{1}{(1-t^2)^2} \left( \prod_{i=1}^2 \oint_{|z^{(2)}_i|=1}  \frac{d z^{(2)}_i}{2 \pi i z^{(2)}_i} \right) \prod_{i\neq j} \left( 1-\frac{z^{(2)}_i}{z^{(2)}_j} \right) /\left( 1- t^2 \frac{z^{(2)}_i}{z^{(2)}_j} \right)  \nn \\
&= \frac{1}{(1-t^2)(1-t^4)}~.
\end{align}
Let us write (as in (A.2) of \cite{Cremonesi:2013lqa}):
\begin{align}
P_{U(2)} (t; m_1, m_2) = \begin{cases} \frac{1}{(1-t^2)^{2}} &\quad m_1 \neq m_2 \\  \frac{1}{(1-t^2)(1-t^4)} &\quad m_1=m_2 \end{cases}~.
\end{align}
Therefore, the Coulomb branch index given in \eqref{CBI1} can be written as
\begin{align}
&g^{-1}(\widetilde{x}=t^2,x=0) \mathcal{I}_C (t; \widetilde{m}^{(-1)}) \nn \\
&= \sum_{m^{(0)} \in \BZ} ~ \sum_{m^{(1)} \in \BZ} ~\sum_{m^{(2)}_1, m^{(2)}_2  \in \BZ}~ \sum_{m^{(3)} \in \BZ} \frac{1}{W(m^{(2)}_1, m^{(2)}_2)} \times \nn \\
& \qquad t^{|m^{(0)} - \widetilde{m}^{(-1)}|+\left( \sum_{i=1}^2 |m^{(0)} - \widetilde{m}^{(2)}_i|+|m^{(1)} - \widetilde{m}^{(2)}_i|+|2m^{(2)}_i-m^{(3)}| \right)-2|m^{(2)}_1-m^{(2)}_2|} \times \nn \\
& \qquad \frac{1}{(1-t^2)^3} P_{U(2)} (t; m_1,m_2)~, \qquad W(m^{(2)}_1, m^{(2)}_2) = \begin{cases} 1 &\quad m^{(2)}_1=m^{(2)}_2 \\  2! &\quad m^{(2)}_1 \neq m^{(2)}_2 \end{cases} \nn \\
&= \sum_{m^{(0)} \in \BZ} ~ \sum_{m^{(1)} \in \BZ} ~\sum_{m^{(2)}_1 \geq m^{(2)}_2  >  -\infty}~ \sum_{m^{(3)} \in \BZ} 
 t^{|m^{(0)} - \widetilde{m}^{(-1)}|+\left( \sum_{i=1}^2 |m^{(0)} - \widetilde{m}^{(2)}_i|+|m^{(1)} - \widetilde{m}^{(2)}_i|+|2m^{(2)}_i-m^{(3)}| \right)} \times \nn \\
&\qquad t^{-2|m^{(2)}_1-m^{(2)}_2|}\frac{1}{(1-t^2)^3} P_{U(2)} (t; m_1,m_2)~.
\end{align}
Upon setting $\widetilde{m}^{(-1)}=0$, the RHS is precisely the Coulomb branch formula presented in \cite{Cremonesi:2014xha} that gives rise to the Hilbert series of one $B_3$ instanton on $\BC^2$:
\begin{align} \label{HSSO7}
{\CI}_C (t; \widetilde{m}^{(-1)}=0) = \frac{1}{(1-t)^2} \times \sum_{p=0}^\infty \dim \; [0,p,0]_{SO(7)} t^{2p}~,
\end{align}
which implies that 
\begin{equation}
g(\widetilde{x},x)\vert_{x=0}=1.
\end{equation}

\section{$\CN=2$ index of the dual of $Sp(1)$ theory with $SO(2N+1)$ flavor symmetry and Chern-Simons level $\kappa=1/2$}\label{Sec:N2Index}
In this final section we shall define the superconformal index of the complete anomaly-free framed affine $B_N$ quiver theory which we have constructed as a dual theory to the $Sp(1)$ theory with $SO(2N+1)$ flavor group and the Chern-Simons term.

The $Sp(1)$ theory in question, and its mirror dual enjoy $\CN=3$ supersymmetry, therefore one should compute the 3d $\CN=2$ index for those theories. Recall the definition of the index on $S^2\times S^1$
\begin{equation}
\mathcal{I} = Tr (-1)^F e^{\beta H} x^{\Delta +j_3} \prod_a t_a^{F_a}, \quad H= \{Q, Q^{\dagger}\}=\Delta - R -j_3\,,
\end{equation}
where $\Delta$ is the energy, $R$ is the R-charge, $j_3$ is the third component of the angular momentum rotating $S^2$, the $F_a$ run over the global flavor symmetry generators. One can obtain the $\CN=2$ index from the $\CN=4$ index by simply setting $\tti=1$ and $x=\tq^{1/2}$ (see previous section). Alternatively, one can use
formulae (2.12) or (2.14) in \cite{Kapustin:2011jm} with the difference that we take the discrete parameters $m$ ($s$ in \cite{Kapustin:2011jm})--which parametrize the GNO charge of the monopole configuration of the gauge field-- to be integers as opposed half-integers.

Recall that the partition function analysis gives the following result for the dual of an $Sp(1)$ theory with $G_f=SO(2N+1)$ and Chern-Simons level $\kappa$.
\begin{equation}
\CZ_{\text{dual}}= \CZ[\widehat{B}_N]+ {\CZ}[\widehat{B}'_N].
\end{equation}
In $\widehat{B}_N$, the double arrow matter is charged under $U(1)_N \times U(2)_{N-1}$ while in the $\widehat{B}'_N$ theory the double arrow matter is charged under $U(1)_N \times SU(2)_{N-1}$ but not under the $U(1)$ subgroup of $U(2)_{N-1}$. This suggests a formula for the $\CN=2$ index of the dual including the Chern-Simons coupling. 

In particular, for theory with $N=3$ we have
\begin{equation}
\begin{split}
\boxed{\mathcal{I}_{dual}(x; k)=f(x,\widetilde{\kappa})\mathcal{I}_{[\widehat{B}_3]}(x) +  g(x,\widetilde{\kappa}) {\mathcal{I}}_{[\widehat{B}'_3]}(x; \widetilde{\kappa}).} \label{dualindex}
\end{split}
\end{equation}
where $\frac{g(x,\widetilde{\kappa})}{f(x,\widetilde{\kappa})} =-1$ and $f(x,\widetilde{\kappa})$ is some arbitrary function of its arguments in agreement with the relative sign of the two contributions to the partition function in \eqref{eq:DualPfSummary}.

The function $\mathcal{I}_{[\widehat{B}'_3]}(x; \widetilde{m}^{(-1)})$ is simply
\begin{empheq}[box=\fbox]{gather} 
\mathcal{I}_{[\widehat{B}_3]}(x;\widetilde{m}^{(-1)})=\sum_{\{m^{(\alpha)}\}} \oint_{|z^{(\alpha)}_i|=1} \frac{dz^{(0)}}{2\pi i z^{(0)}} \frac{dz^{(1)}}{2\pi i z^{(1)}} \frac{dz^{(3)}}{2\pi i z^{(3)}} \frac{1}{W(m^{(2)})} \prod_{i=1,2} \frac{dz^{(2)}_i}{2\pi i z^{(2)}_i} \nonumber \\
\times  \mathcal{I}_\text{fund}^{(m^{(0)}, \widetilde{m}^{(-1)})} (z^{(0)}, \widetilde{z}^{(0)})\mathcal{I}_\text{bifund}^{(m^{(0)}, m^{(2)})} (z^{(0)}, z^{(2)})\mathcal{I}_\text{bifund}^{(m^{(1)}, m^{(2)})} (z^{(1)}, z^{(2)}) \mathcal{I}^{(m^{(3)}, m^{(2)})} _{\text{nsl}}(z^{(3)},z^{(2)}) \nonumber \\
\times  \mathcal{I}_V^{(m^{(0)})}(z^{(0)}) \mathcal{I}_V^{(m^{(1)})}(z^{(1)}) \mathcal{I}_V^{(m^{(2)})}(z^{(2)}) \mathcal{I}_V^{(m^{(3)})}(z^{(3)}). \label{IB3}
\end{empheq}

Similarly $\widetilde{\mathcal{I}}^{(n)}_{[\widehat{B}'_3]}(x, \widetilde{\kappa}; \widetilde{m}^{(-1)},w,a)$ can be written as
\begin{empheq}[box=\fbox]{gather} 
\widetilde{\mathcal{I}}^{(n)}_{[\widehat{B}'_3]}(x,k; \widetilde{m}^{(-1)},w,a)=\sum_{\{m^{(\alpha)}\}} \oint_{|z^{(\alpha)}_i|=1} \frac{dz^{(0)}}{2\pi i z^{(0)}} \frac{dz^{(1)}}{2\pi i z^{(1)}} \frac{dz^{(3)}}{2\pi i z^{(3)}} \frac{1}{W(m^{(2)})} \prod_{i=1,2} \frac{dz^{(2)}_i}{2\pi i z^{(2)}_i} \nonumber\\
\times \mathcal{I}_\text{fund}^{(m^{(0)}, \widetilde{m}^{(-1)})} (z^{(0)}, \widetilde{z}^{(0)}) \mathcal{I}_\text{bifund}^{(m^{(0)}, m^{(2)})} (z^{(0)}, z^{(2)})\mathcal{I}_\text{bifund}^{(m^{(1)}, m^{(2)})} (z^{(1)}, z^{(2)}) \mathcal{I}_V^{(m^{(2)})}(z^{(2)}) \nonumber \\
\times  \mathcal{I}_V^{(m^{(0)})}(z^{(0)}) \mathcal{I}_V^{(m^{(1)})}(z^{(1)})  \mathcal{I}_V^{(m^{(3)})}(z^{(3)}) \nonumber  \\
\times \widetilde{\mathcal{I}}^{(m^{(3)}, m^{(2)})} _{\text{nsl}}(z^{(3)},z^{(2)}) \times \mathcal{I}_\text{CS}(z^{(3)},m^{(3)},\widetilde{\kappa}) \times \mathcal{I}_\text{FI}(z^{(3)},m^{(3)},w,a). \label{IB3p}
\end{empheq}

The various functions appearing in the integrand of $\mathcal{I}_{[\widehat{B}_3]}(x; \widetilde{m}^{(-1)})$ are defined as follows:
\begin{align} 
&\mathcal{I}_\text{fund}^{(m^{(0)}, \widetilde{m}^{(-1)})} (x, z^{(0)}, \widetilde{z}^{(0)})=(x)^{\frac{1}{2}|m^{(0)}-\widetilde{m}^{(-1)}|} \frac{(x^{3/2+|m^{(0)}-\widetilde{m}^{(-1)}|} (z^{(0)}/\widetilde{z}^{(0)})^{\pm 1};x^2)}{(x^{1/2+2|m^{(0)}-\widetilde{m}^{(-1)}|} (z^{(0)}/\widetilde{z}^{(0)})^{\pm 1};x^2)}, \label{sci_fund4}\\
& \mathcal{I}_\text{bifund}^{(m^{(\alpha)}, {m}^{(2)})} (x,z^{(\alpha)}, {z}^{(2)})=\prod_{i=1,2}(x)^{\frac{1}{2} |m^{(\alpha)}-{m}_i^{(2)}|} \frac{(x^{3/2+|m^{(\alpha)}-{m}_i^{(2)}|} (z^{(\alpha)}/{z}_i^{(2)})^{\pm 1};x^2)}{(x^{1/2+|m^{(\alpha)}-{m}_i^{(2)}|} (z^{(\alpha)}/{z}_i^{(2)})^{\pm 1};x^2)},\label{sci_bifund4}\\
&\mathcal{I}_V^{(m^{(i)})}(x,z^{(i)}) =1\,, \quad i=0,1,3\label{sci_abV},\\
&\mathcal{I}_V^{(m^{(2)})}(z^{(2)}) = \prod_{i \neq j} (x)^{-\frac{1}{2} |m_i^{(2)}-{m}_j^{(2)}|} (1-x^{|m^{(2)}_i -m_j^{(2)}|}z_i^{(2)}/{z}_j^{(2)} ), \label{sci_nonabV4}\\
&\mathcal{I}^{(m^{(3)}, m^{(2)})} _{\text{nsl}}(z^{(3)},z^{(2)})= \prod_{i=1,2}(x)^{\frac{1}{2} |m^{(3)}-2{m}_i^{(2)}|} \frac{(x^{3/2+|m^{(3)}-2{m}_i^{(2)}|} (z^{(3)}/({z}_i^{(2)})^2)^{\pm 1};x^2)}{(x^{1/2+|m^{(\alpha)}-2{m}_i^{(2)}|} (z^{(3)}/({z}_i^{(2)})^2)^{\pm 1};x^2)}. \label{sci_nsl14}
\end{align}


Note that we do not have any Chern-Simons term in $\mathcal{I}_{[\widehat{B}_3]}(\tq,\tti; \widetilde{m}^{(-1)})$ or any FI term (coupling with the background $U(1)_J$ for any of the gauge groups). \\

The contributions of the fundamental/bifundamental matter and the different gauge groups in $\widetilde{\mathcal{I}}^{(n)}_{[\widehat{B}'_3]}(x, \widetilde{\kappa}; \widetilde{m}^{(-1)},w,a)$ are given by \eqref{sci_fund4}--\eqref{sci_nonabV4} as before while the contribution of the double arrow, the Chern-Simons and FI terms for the node with Dynkin label ``3" in the $\widehat{B}'_3$ quiver are
\begin{align}
& \widetilde{\mathcal{I}}^{(m^{(3)}, m^{(2)})} _{\text{nsl}}(z^{(3)},z^{(2)}) =\prod_{s=\pm 1}x^{\frac{1}{2} |m^{(3)}-s({m}_1^{(2)}-m_2^{(2)})|} \frac{\left(x^{3/2+|m^{(3)}-s({m}_1^{(2)}-m_2^{(2)})|} \left[z^{(3)}\left(\frac{{z}_2^{(2)}}{{z}_1^{(2)}}\right)^s\right]^{\pm 1};x^2\right)}{\left(x^{1/2+|m^{(3)}-s({m}_1^{(2)}-m_2^{(2)})|} \left[z^{(3)}\left(\frac{{z}_2^{(2)}}{{z}_1^{(2)}}\right)^s\right]^{\pm 1};x^2\right)},\\
& \mathcal{I}_\text{CS}(z^{(3)},m^{(3)},\widetilde{\kappa})=(z^{(3)})^{\widetilde{\kappa} m^{(3)}},  \\
&\mathcal{I}_\text{FI}(z^{(3)},m^{(3)},w,a)= (z^{(3)})^{2a} w^{2m^{(3)}}\,,
 \label{Inslprime}
\end{align}
where we recall from \eref{ZBN} that the Chern-Simons level for the right-most node in the $\widehat{B}$ quiver is $\widetilde{\kappa}=-1$.

\section*{Acknowledgements}
We are much grateful to Kavli Institute for Theoretical Physics at University of California Santa Barbara, where this project has started during program``New Methods in Nonperturbative Quantum Field Theory'' in 2014. Also we would like to thank Simons Center for Geometry and Physics at Stony Brook University and especially to Cumrun Vafa and Martin Ro$\check{\text{c}}$ek for organizing the 2014, 2015 Summer Workshops.
PK would also like to thank the Theoretical Physics group at Imperial College in London and W. Fine Institute for Theoretical Physics at University of Minnesota for kind hospitality during his visit, where part of his work was done. In addition PK thanks Jaume Gomis and Viktor Mikhailov for fruitful discussions. 
The research of PK was supported in part by the Perimeter Institute for Theoretical Physics. Research at Perimeter Institute is supported by the Government of Canada through Industry Canada and by the Province of Ontario through the Ministry of Economic Development and Innovation. 
NM thanks the CERN visitor programme from October to November 2016, during which this work is finalized.   He is also supported in part by the INFN and by the ERC Starting Grant 637844-HBQFTNCER.
This work was performed in part at the Aspen Center for Physics, which is supported by National Science Foundation grant PHY-1066293.

\appendix
\section{Cauchy Identity and Fourier Transform}\label{Sec:CauchyFourier}
Starting from \eqref{pfBtriv-M=1} we wish to use the Cauchy identity
\begin{equation}
\frac{1}{ \cosh{\pi (s^{1}+m)}\cosh{\pi (s^{2}+m')}}-\frac{1}{ \cosh{\pi (s^{2}+m)}\cosh{\pi (s^{1}+m')}}=\frac{\sh{(s^{1}-s^{2})}\sh{(m-m')}}{\prod^{2}_{p=1} \ch{(s^{p}+m)}\ch{(s^{p}+m'})}\,,
\end{equation}

For this purpose, we first introduce a delta function into the integration, and replace $s$ by $s^1$ and $s^2$. The numerator is split to get,
\begin{equation}
\begin{split}
\CZ_A&=\int \frac{d^{2} s}{2!} \left( \frac{\delta(s^{1}+s^{2}) \sh{(s^{1}-s^{2})}e^{2i \pi \kappa (s^{1})^2 }}{\ch{s^{1}}}\right) \frac{1}{\prod^{N-2}_{a=1}\prod^{2}_{p=1}\ch{(s^{p}+m_a)}}\\
&\times \left(\frac{\sh{(s^{1}-s^{2})}}{\prod^{2}_{p=1} \ch{(s^{p}+m_{N-1})}\ch{(s^{p}+m_N})} \right)\times \frac{1}{\cosh{\pi (s^1+s^2-M_{as})}}\\
\end{split}
\end{equation}
Next we introduce a permutation group in 2 elements $S_2$ and denote a permutation by an element $\rho\in S_2$.
The equation is now ready for applying the identity and we replace to get
\begin{equation}
\begin{split}
&=\int \frac{d^{2} s}{2!} \left(\frac{\delta(s^1+s^{2})e^{2 i \pi \kappa (s^1)^2} \sh{(s^1-s^{2})}}{\ch{s^1}}\right) \frac{1}{\prod^{N-2}_{a=1}\prod^{2}_{p=1}\ch{(s^p+m_a)}}\\
& \times \left(\sum _{\rho\in S_2} (-1)^{\rho} \frac{(\sh{(m_{N-1}-m_{N})})^{-1}}{ \cosh{\pi (s^{\rho(1)}+m_N)}\cosh{\pi (s^{\rho(2)}+m_{N-1})}} \right) \times \frac{1}{\cosh{\pi (s^1+s^2-M_{as})}}
\end{split}
\end{equation}
Here is a shorter way of writing the identity:
\begin{equation}
\sum _{\rho\in S_2} (-1)^{\rho} \frac{1}{ \cosh{\pi (s^{\rho(1)}+m_N)}\cosh{\pi (s^{\rho(2)}+m_{N-1})}}=\frac{\sh{(s^{(1)}-s^{(2)})}\sh{(m_{N-1}-m_{N})}}{\prod^{2}_{p=1} \ch{(s^p+m_{N-1})}\ch{(s^{p}+m_N})}\,,
\end{equation}
where $(-1)^{\rho}$ is the sign of the permutation $\rho$. 

In the next step, we introduce a set of auxiliary variables $s_\beta^p,\, \beta=0,\dots, N-2,\, p=1,2$ in the following way
\begin{equation}
\begin{split}
\CZ_A&= \int \prod^{N-2}_{\beta=0}\frac{d^{2} s_\beta}{2!} \left(\frac{\delta(s^1_0+s^{2}_0)e^{2 i \pi \kappa (s^1_0)^2}\sh{(s^1_0-s^{2}_0)}}{\ch{s^1_0}}\right)  \prod^{N-3}_{\beta=0} \frac{\prod^{2}_{p=1}\delta(s^p_\beta-s^{p}_{\beta +1})}{\prod^{2}_{p=1}\ch{(s^p_{\beta+1}+m_{\beta+1})}}\\
&\times  \left(\sum _{\rho} (-1)^{\rho} \frac{(\sh{(m_{N-1}-m_{N})})^{-1}}{ \cosh{\pi (s^{\rho(1)}_{N-2}+m_N)}\cosh{\pi (s^{\rho(2)}_{N-2}+m_{N-1})}} \right)\times \frac{1}{\cosh{\pi (s^1_{N-2}+s^2_{N-2}-M_{as})}}\\
\end{split}
\end{equation}
S-duality is implemented by rewriting the integral in terms of Fourier transform/dual variables $u_0,\dots, u_{N-2}$ and $\tau_{1}$. Appropriately anti-symmetrizing the integrand, we obtain
\begin{equation}
\begin{split}
\CZ_A=&\int \prod^{N-2}_{\beta=0} d^{2} s_{\beta} d^{2} u_{\beta}  d \tau_{1} \left(\frac{\delta(s^1_0+s^{2}_0)e^{2 i \pi \kappa (s^1_0)^2}\sh{(2s^1_0)}}{\ch{s^1_0}} \right) \\
&\times \prod^{N-3}_{\beta=0}  \left(\sum_{\rho_{\beta}} (-1)^{\rho_{\beta}}\prod^2_{p=1} \frac{e^{2\pi i u^p_{\beta}(s^p_{\beta} - s_{\beta+1}^{\rho_{\beta}(p)})}}{\ch{(s^p_{\beta+1} +m_{\beta+1})}}\right)\\
&\times \left(\sum_{\rho} (-1)^{\rho}  \frac{e^{2\pi i u^1_{N-2}(s^{\rho(1)}_{N-2}+m_{N-1})}e^{2\pi i u^{2}_{N-2}(s^{\rho(2)}_{N-2}+m_{N})} e^{2\pi i \tau_1 (s^{\rho(1)}_{N-2}+s^{\rho(2)}_{N-2}-M_{as})}}{\ch{u^1_{N-2}} \ch{u^{2}_{N-2}} \ch{\tau_1}\sh{(m_{N-1}-m_{N})}} \right)
\end{split}
\end{equation}

In the next step, we need to integrate over the variables $\{s^i_{\beta}\}$ to obtain the dual partition function after rearranging terms in the integrand in the following fashion.
\begin{equation}
\begin{split}
&\CZ_A=\int \prod^{N-2}_{\beta=0} d^{2} s_{\beta} d^{2} u_{\beta} d\tau_{1}\left(\frac{\delta(s^1_0+s^{2}_0)e^{2 i \pi \kappa (s^1_0)^2}\sh{(2s^1_0)}\prod_p e^{2\pi i u^p_0 s^p_0} e^{2\pi i m_1 u^p_0}}{\ch{s^1_0}} \right)\\
&\times \prod^{N-3}_{\beta=1} \left(\sum_{\rho_{\beta-1}} (-1)^{\rho_{\beta-1}}  \prod^2_{p=1} \frac{e^{2\pi i (s^p_{\beta} +m_{\beta})(u^p_{\beta} - u^{\rho^{-1}_{\beta -1} (p)}_{\beta -1})}}{\ch{(s^p_{\beta} +m_{\beta})}}  \prod_p e^{-2\pi i m_{\beta} (u^p_{\beta} -u^p_{\beta-1})}\right) \prod_p e^{-2\pi i m_1 u^p_0}\\
&\times\sum_{\rho, \rho_{N-3}} (-1)^{\rho+\rho_{N-3}} \\
& \frac{\exp\left[(2\pi i (s^{\rho(1)}_{N-2}+m_{N-2})(u^1_{N-2} +\tau_1 - u^{\rho\circ \rho^{-1}_{N-3}(1)}_{N-3})\right] \exp\left[2\pi i (s^{\rho(2)}_{N-2}+m_{N-2})(u^{2}_{N-2} +\tau_1 - u^{\rho\circ \rho^{-1}_{N-3}(2)}_{N-3})\right]}{\prod_p \ch{u^p_{N-2}}\ch{(s^p_{N-2}+m_{N-2})}\ch{\tau_1}\sh{(m_{N-1}-m_N)}}\\
&\times \prod_p e^{2\pi iu^p_{N-3} m_{N-2}}  \times e^{2\pi i u^1_{N-2} (m_{N-1}- m_{N-2})} e^{2\pi i u^{2}_{N-2} (m_N- m_{N-2})} e^{-2\pi i \tau_1 (M+ 2m_{N-2})}\\
&\equiv \int  X(u_0,s_0) \; Y (u_0, s_1,u_1,\ldots, s_{N-3},u_{N-3})\; Z(u_{N-3}, s_{N-2}, u_{N-2}, \tau_1)\,,
\end{split}
\label{eq:ZAXYZ}
\end{equation}
where $X$ denotes the contribution from the first line, $Y$ from the second and third lines, and $Z$ shows the last line. The $X(u_0,s_0)$ contains the information about the double bond.
Performing the integrals over $\{s_1,s_2,\ldots,s_{N-2}\}$ is straightforward and explained in appendix \S\ref{YZ}. The integral over the $s_0$--dependent piece yields the contribution of the double bond to the dual partition function and we proceed to compute that next.

We are ready to complete the desired partition function of the new $\widehat{B}_{N}$-type quiver gauge theory.
Let us rewrite the partition function after  partial integrations over $Y$ and $Z$ from \eqref{eq:ZAXYZ} and redefining the variable $u^p_{N-2} \to u^p_{N-2} -\tau_1$:
\begin{equation}
\begin{split}\label{Sdual1}
&\CZ_A (\vec{m},\kappa)\\
&= \int \prod^{N-2}_{\beta=1} \frac{d^{2} u_{\beta}}{2!} \prod^2_{\alpha=1} d\tau_{\alpha} \int \frac{d^2u_0}{2!} \frac{d^{2} s_{0}}{2!}\prod_p e^{2\pi i u^p_0 s^p_0} e^{2\pi i m_1 u^p_0} \\
&\times \frac{\delta(s^1_0+s^{2}_0)e^{2 i \pi \kappa (s^1_0)^2}\sh{(2s^1_0)}\sh{(u^1_{0} - u^2_{0})}}{\ch{s^1_0}\prod^2_{p,l=1} \ch{(u^p_{0} - u^l_{1})}}\\
&\times  \frac{\prod^{N-2}_{\beta=1} \sinh^2{\pi (u^1_{\beta} - u^2_{\beta})}}{\prod^{N-3}_{\beta=1}\prod^2_{p,l=1} \ch{(u^p_{\beta} - u^l_{\beta+1})}} \times \prod^{N-2}_{\beta=1} \prod^2_{p=1} e^{2\pi i \widetilde{\eta}_{\beta} u^p_{\beta}} \\
& \times \frac{ -i e^{2\pi i \eta_1 \tau_1 }e^{2\pi i \eta_2 \tau_2 }}{\prod_{p} \ch{(u^p_{N-2}-\tau_1)}\ch{(u^p_{N-2}-\tau_2)} \ch{\tau_1}}\\
&= \mathcal{Z}_B.
\end{split}
\end{equation}
where $\mathcal{Z}_B$ is the dual partition function and the various FI parameters will be explicitly given as functions of masses in the next section. In the above integrand the last two lines correspond to the known contribution of the left ($D$-type) tail of the quiver, whereas the first two give a contribution of the double arrow of the $\widehat{B}_{N}$ quiver theory.

Labeling the contribution of the double arrow as $\mathcal{Z}_{\text{nsl}}$\footnote{\text{nsl} for non-simply laced}, after integrating over $s_0^1, s_0^2$ and $u_0^1$, the dual partition function can be written as
\begin{equation}
\begin{split}\label{Sdual2}
&\mathcal{Z}_B:= \int \prod^{N-2}_{\beta=1} \frac{d^{2} u_{\beta}}{2!} \prod^2_{\alpha=1} d\tau_{\alpha} \int du_0^2\, \mathcal{Z}_{\text{nsl}}(u^2_0,u^l_1;\kappa,m_1) \\
& \times  \frac{\prod^{N-2}_{\beta=1} \sinh^2{\pi (u^1_{\beta} - u^2_{\beta})} \prod^{N-2}_{\beta=1} \prod^2_{p=1} e^{2\pi i \widetilde{\eta}_{\beta} u^p_{\beta}}}{\prod^{N-3}_{\beta=1}\prod^2_{p,l=1} \ch{(u^p_{\beta} - u^l_{\beta+1})}} \\
&\times \frac{ e^{2\pi i \eta_1 \tau_1 }e^{2\pi i \eta_2 \tau_2 }}{\prod_{p} \ch{(u^p_{N-2}-\tau_1)}\ch{(u^p_{N-2}-\tau_2)} \ch{\tau_1}}\,.
\end{split}
\end{equation}
For simplifying the computation, we set $m_1=0$\footnote{$m_1 \neq 0$ case does not seem to lead to an easy dual interpretation -- for example, it breaks the $U(2)$ gauge symmetry of the node parametrized by $\{u^l_1\}$} and after a rather tedious computation detailed in the \appref{Sec:Znsl} we get
\begin{equation}
\begin{split}\label{nsl0}
&\int du_0^2 \; \mathcal{Z}_{\text{nsl}}(u_0^2,u^l_1;\kappa,m_1=0) \\
=&-i\int  \frac{d^{2} u_{0}}{2!}  \frac{d^{2} s_{0}}{2!}\left(\frac{\delta(s^1_0+s^{2}_0)e^{2 i \pi \kappa (s^1_0)^2}\sh{(2s^1_0)}\sh{(u^1_{0} - u^2_{0})}}{\ch{s^1_0}\prod^2_{p,l=1} \ch{(u^p_{0} - u^l_{1})}} \right) \prod_p e^{2\pi i u^p_0 s^p_0} \\
=& i e^{-i\kappa\pi/2} \int  d u^2_{0} ds e^{2 i \pi \kappa s^2} \frac{e^{\pi s}}{\sh{s}}\sh{2\kappa s}\left(\frac{e^{2i\pi (u^2_1+u^1_1-u^2_0)s}}{\ch{(u^2_0-2 u^1_1)}\ch{(u^2_0-2 u^2_1)}} \right) \\
&+ {e^{-i\kappa\pi/2}} \int d u^2_{0} \frac{1}{\ch{(u^2_0-2 u^1_1)}\ch{(u^2_0-2 u^2_1)}}.
\end{split}
\end{equation}
Note that if $\kappa=0$ then the first integral vanishes so that the second term can be identified with the dual partition function of the anomalous $Sp(k)$ ADHM theory. However, we are interested in nonzero Chern-Simons level, namely $\kappa=1/2$, which makes the theory $A$ anomaly free. 

One can massage the first integral in the above formula into a more convenient form by completing the integration over $s$ and shifting the integration variable $u^2_0 \to u^2_0+u^2_1+u^1_1$
\begin{equation}
\begin{split}\label{nsl1}
& ie^{-i\kappa \pi/2} \int  d u^2_{0} ds\, e^{2 i \pi \kappa s^2} e^{\pi s}\frac{\sinh{2\pi\kappa s}}{\sh{s}}\,\frac{e^{2i\pi (u^2_1+u^1_1-u^2_0)s}}{\ch{(u^2_0-2 u^1_1)}\ch{(u^2_0-2 u^2_1)}} \\
\overset{\kappa=1/2}=& i {e^{i\pi/4}}\int  d u^2_{0}\,  e^{-i \pi (u^2_0)^2}  \frac{e^{\pi u^2_0}}{\ch{(u^2_0-u^1_1+u^2_1)}\ch{(u^2_0- u^2_1+u^1_1)}}\,.
\end{split}
\end{equation}

Now let us put all the pieces together to write the dual partition function (after renaming the integration variable $u \to u_0$):
\begin{equation}
\begin{split}
&\CZ_{B}= \CZ_A[k;m_1=0, m_2,\ldots,m_N]\\
=& \int du_0  \prod^{N-2}_{\beta=1} \frac{d^{2} u_{\beta}}{2!} \prod^2_{\alpha=1} d\tau_{\alpha}  \mathcal{Z}_{\text{nsl}}(u_0,u^l_1;\kappa,m_1=0)\\
&\times  \frac{\prod^{N-2}_{\beta=1} \sinh^2{\pi (u^1_{\beta} - u^2_{\beta})}}{\prod^{N-3}_{\beta=1}\prod^2_{p,l=1} \ch{(u^p_{\beta} - u^l_{\beta+1})}} \times \prod^{N-2}_{\beta=1} \prod^2_{p=1} e^{2\pi i \widetilde{\eta}_{\beta} u^p_{\beta}} \\
& \times \frac{ e^{2\pi i \eta_1 \tau_1 }e^{2\pi i \eta_2 \tau_2 }}{\prod_{p} \ch{(u^p_{N-2}-\tau_1)}\ch{(u^p_{N-2}-\tau_2)} \ch{\tau_1}}\,,
\end{split}
\end{equation} 
where the function $ \mathcal{Z}_{\text{nsl}}(u_0,u^l_1;\kappa,m_1=0)$ can be computed from \eqref{nsl0} and \eqref{nsl1} 
\begin{equation}
\begin{split}\label{Z-nsl}
\int du_0 \mathcal{Z}_{\text{nsl}}(u_0,u^l_1;\kappa,m_1=0)=&  i {e^{i\pi/4}}\int  d u_{0}   \left(\frac{e^{-i \pi (u_0)^2}e^{\pi u_0}}{\ch{(u_0-u^1_1+u^2_1)}\ch{(u_0- u^2_1+u^1_1)}} \right)\\
& + {e^{-i\pi/4}} \int d u_{0}  \left(\frac{1}{\ch{(u_0-2 u^1_1)}\ch{(u_0-2 u^2_1)}} \right) .
\end{split}
\end{equation} 

Let us now label the Cartan of the nodes in direct correspondence of their Dynkin labels of the $\widehat{B}_{N}$ quiver diagram (see \figref{fig:quiverBN}) 
\begin{equation}
u_0 \to u_N\,,\, u_\beta^a \to u_{N-\beta}^a\,, \tau_2 \to u_1,\, \tau_1 \to u_0,
\end{equation}
where $\beta=1,\ldots,N-2$.
Then the function for Chern-Simons level $\kappa=1/2$ $\mathcal{Z}_{\text{nsl}}$ becomes
\begin{empheq}[box=\fbox]{gather}\label{eq:BoxedFnsl}
\mathcal{Z}_{\text{nsl}}(u_N,u^l_{N-1})={e^{-i\pi/4}} \left(\frac{1}{\ch{(u_N-2 u^1_{N-1})}\ch{(u_N-2 u^2_{N-1})}} \right)\\      \nonumber
 +i {e^{i\pi/4}} \left(\frac{e^{-i \pi (u_N)^2} e^{\pi u_N}}{\ch{(u_N-u^1_{N-1}+u^2_{N-1})}\ch{(u_N- u^2_{N-1}+u^1_{N-1})}} \right)\\ \nonumber
 =: {e^{-i\pi/4}} F^{(1)}_{\text{nsl}}(u_N,u^l_{N-1}) +  i {e^{i\pi/4}} e^{-i \pi (u_N)^2} e^{\pi u_N} F^{(2)}_{\text{nsl}}(u_N,u^l_{N-1}),
\end{empheq}
where the functions $F^{(1)}_{\text{nsl}}(u_N,u^l_{N-1})$ and $F^{(2)}_{\text{nsl}}(u_N,u^l_{N-1})$ are:
\begin{equation}
\begin{split}
&F^{(1)}_{\text{nsl}}(u_N,u^l_{N-1}) = \frac{1}{\ch{(u_N-2 u^1_{N-1})}\ch{(u_N-2 u^2_{N-1})}}\,,\\
& F^{(2)}_{\text{nsl}}(u_N,u^l_{N-1})= \frac{1}{\ch{(u_N-u^1_{N-1}+u^2_{N-1})}\ch{(u_N- u^2_{N-1}+u^1_{N-1})}}\,.
\label{eq:FnsldefA}
\end{split}
\end{equation}

\section{Computation of $Y$ and $Z$}\label{YZ}
First consider the partial integration of $Y$.
\begin{equation}
\begin{split}
& \int \prod^{N-3}_{\beta=1} d^{2} s_{\beta} Y (u_0, s_1,u_1,\ldots, s_{N-3},u_{N-3})\\
&=\prod^{N-3}_{\beta=1}\left(\sum_{\rho_{\beta-1}} (-1)^{\rho_{\beta-1}}  \prod_p \frac{e^{-2\pi i m_{\beta} (u^p_{\beta} -u^p_{\beta-1})}}{\ch{(u^p_{\beta} - u^{\rho^{-1}_{\beta -1} (p)}_{\beta -1})}} \right) \prod_p e^{-2\pi i m_1 u^p_0}\\
&= \prod_p e^{-2\pi i m_1 u^p_0} \prod^{N-3}_{\beta=1} \left(\frac{\sh{(u^1_{\beta-1} - u^2_{\beta-1})} \sh{(u^1_{\beta} - u^2_{\beta})}}{\prod^2_{p,l=1} \ch{(u^p_{\beta-1} - u^l_{\beta})}} \times \prod^2_{p=1} e^{-2\pi i m_{\beta} (u^p_{\beta} -u^p_{\beta-1})}\right).
\end{split}
\end{equation}
Now consider the partial integration of $Z$.
\begin{equation}
\begin{split}
&\int  d^{2} s_{N-2} Z(u_{N-3}, s_{N-2}, u_{N-2}, \tau_1)\\
=-i&\int d \tau_2 \left(\frac{\sh{(u^1_{N-3} - u^2_{N-3})} \sinh^2{\pi (u^1_{N-2} - u^2_{N-2})}}{\prod^2_{p,l=1} \ch{(u^p_{N-2} - u^l_{N-3})}}\right)\\
&\times \left(\frac{ e^{-2\pi i \tau_1 (M_{as} +m_{N}+m_{N-1})}e^{2\pi i \tau_2 (m_{N}- m_{N-1})}}{\prod_{p} \ch{(u^p_{N-2}-\tau_1)}\ch{(u^p_{N-2}-\tau_2)} \ch{\tau_1}}\right)\\
&\times \left(\prod_p e^{2\pi iu^p_{N-3} m_{N-2}}\prod_p e^{2\pi i u^p_{N-2} (m_{N-1}- m_{N-2})} \right)
\end{split}
\end{equation}
The new auxiliary variable $\tau_2$ which labels the Cartan of one of the boundary $U(1)$ nodes in the dual theory comes from the following identity which has been used to obtain the above result.
\begin{equation}
\begin{split}
&\frac{i}{\sh{\eta}} \left( e^{2\pi i \eta {u}^2_{N-2} }\right)  \left(2\sinh{\pi(u^1_{N-2} -u^{2}_{N-2})} \right)^{-1} \Big\vert_{\{u^p_{N-2}\}}=\int d \tau_2  \frac{e^{2\pi i \eta \tau_2}}{\prod_{i,p}\cosh{\pi (\tau_2 -u^p_{N-2})} } 
\end{split}\label{integrating in node_1}
\end{equation}
where $\{u^p_{N-2}\}$ denotes symmetrization w.r.t. the said variables which requires simply multiplying by some combinatorial factor since the rest of the integrand is symmetric in these variables. Also, in the above formula $\eta =m_{N} -m_{N-1}$. \\

Now, we can read off the FI parameters as functions of various masses; note that we identify the exponents of $e^{2\pi i \tau_{1,2}} $, $e^{2\pi i u^p_{\beta}}$ as the respective FI parameters.
The full dictionary then reads as follows
\begin{equation}
\begin{split}
&\eta_0=-M_{as} -m_{N} - m_{N-1}\,, \\ 
&\eta_\beta=m_{N-\beta+1} - m_{N-\beta }\,,\qquad  \beta=1,\ldots,N-2\,,\\
&\eta_{N-1}=m_2,\\
&\eta_N=0\,,
\end{split}
\label{eq:MirrorMap0}
\end{equation}
with Fayet-Iliopoulos parameters of the framed affine $B_N$ quiver on the left hand sides of the above equations and masses of $SO(2N+1)$ chirals and mass $M_{as}$ of the anti-symmetric $Sp(1)$ matter on the right. It is instructive to redefine the chiral masses as $m_{N-\beta+1} \to m_\beta$ (therefore $m_{N-\beta} \to  m_{\beta+1}$) so that the duality map reflects the structure of simple roots associated with the $B_N$ Dynkin diagram (summarized in \eqref{eq:DictionarymassesfI}) :
\begin{equation}
\begin{split}
&\eta_0=-M_{as} - m_{1} -m_{2}\,, \\ 
&\eta_\beta=m_{\beta} -m_{\beta +1 }\,,\quad  \beta=1,2,3,\ldots,N-2\,,\\
&\eta_{N-1}=m_{N-1},\\
&\eta_N=0\,.
\end{split}
\label{eq:MirrorMap1}
\end{equation}

\section{Computation of $\mathcal{Z}_{\text{nsl}}$}\label{Sec:Znsl}
Recall the formula for the partition function of the $Sp(1)$ Chern-Simons theory with an $SO(2N+1)$ flavor symmetry and a free hypermultiplet obtained in \eqref{Sdual2}
\begin{equation}\label{ZAfullB}
\begin{split}
&\mathcal{Z}_A= \int \prod^{N-2}_{\beta=0} \frac{d^{2} u_{\beta}}{2!} \prod^2_{\alpha=1} d\tau_{\alpha} \frac{d^{2} s_{0}}{2!}\left(\frac{\delta(s^1_0+s^{2}_0)e^{2 i \pi \kappa (s^1_0)^2}\sh{(2s^1_0)}\sh{(u^1_{0} - u^2_{0})}}{\ch{s^1_0}\prod^2_{p,l=1} \ch{(u^p_{0} - u^l_{1})}} \right) \prod_p e^{2\pi i u^p_0 s^p_0} e^{2\pi i m_1 u^p_0}\\
&\times  \left(\frac{\prod^{N-2}_{\beta=1} \sinh^2{\pi (u^1_{\beta} - u^2_{\beta})}}{\prod^{N-3}_{\beta=1}\prod^2_{p,l=1} \ch{(u^p_{\beta} - u^l_{\beta+1})}} \times \prod^{N-2}_{\beta=1} \prod^2_{p=1} e^{2\pi i \widetilde{\eta}_{\beta} u^p_{\beta}}\right) \\
& \times \left(\frac{ -i e^{2\pi i \eta_1 \tau_1 }e^{2\pi i \eta_2 \tau_2 }}{\prod_{p} \ch{(u^p_{N-2}-\tau_1)}\ch{(u^p_{N-2}-\tau_2)} \ch{\tau_1}}\right)\,,
\end{split}
\end{equation}
which is equal to the partition function of the mirror dual theory
\begin{equation}
\begin{split}\label{Sdual2B}
&\mathcal{Z}_B=\int du \prod^{N-2}_{\beta=1} \frac{d^{2} u_{\beta}}{2!} \prod^2_{\alpha=1} d\tau_{\alpha} \mathcal{Z}_{\text{nsl}}(u,u^l_1;\kappa,m_1) \times  \left(\frac{\prod^{N-2}_{\beta=1} \sinh^2{\pi (u^1_{\beta} - u^2_{\beta})} \prod^{N-2}_{\beta=1} \prod^2_{p=1} e^{2\pi i \widetilde{\eta}_{\beta} u^p_{\beta}}}{\prod^{N-3}_{\beta=1}\prod^2_{p,l=1} \ch{(u^p_{\beta} - u^l_{\beta+1})}}  \right) \\
&\times \left(\frac{ e^{2\pi i \eta_1 \tau_1 }e^{2\pi i \eta_2 \tau_2 }}{\prod_{p} \ch{(u^p_{N-2}-\tau_1)}\ch{(u^p_{N-2}-\tau_2)} \ch{\tau_1}}\right)\,,
\end{split}
\end{equation}
and $\mathcal{Z}_{\text{nsl}}$ is given in \eqref{nsl0}. Now, let us manipulate the $u^p_0$-dependent part of $\mathcal{Z}_A$, i.e. the first line of \eqref{ZAfullB}
\begin{equation}
\begin{split}
&\int  \frac{d^{2} u_{0}}{2!}  \frac{d^{2} s_{0}}{2!}\left(\frac{\delta(s^1_0+s^{2}_0)e^{2 i \pi \kappa (s^1_0)^2}\sh{(2s^1_0)}\sh{(u^1_{0} - u^2_{0})}}{\ch{s^1_0}\prod^2_{p,l=1} \ch{(u^p_{0} - u^l_{1})}} \right) \prod_p e^{2\pi i u^p_0 s^p_0} e^{2\pi i m_1 u^p_0}\\
=& \int  \frac{d^{2} u_{0}}{2!} ds \left(\frac{e^{2 i \pi \kappa s^2}\sh{2s}\sh{(u^1_{0} - u^2_{0})}}{\ch{s}\prod^2_{p,l=1} \ch{(u^p_{0} - u^l_{1})}} \right)  e^{2\pi i s(u^1_0-u^2_0)} e^{2\pi i m_1 (u^1_0+u^2_0)}\\
=& \int  \frac{d^{2} u_{0}}{2!} ds \frac{e^{2 i \pi \kappa s^2}\sh{(u^1_{0} - u^2_{0})}}{\ch{s}\prod^2_{p,l=1} \ch{(u^p_{0} - u^l_{1})}} \frac{1}{2}\left(e^{2\pi i s(u^1_0-u^2_0-i)}-e^{2\pi i s(u^1_0-u^2_0+i)}\right)e^{2\pi i m_1 (u^1_0+u^2_0)}\\
=& \int  \frac{d^{2} u_{0}}{2!} ds \frac{e^{2 i \pi \kappa s^2}\sh{(u^1_{0} - u^2_{0})}}{\ch{s}\prod^2_{p,l=1} \ch{(u^p_{0} - u^l_{1})}} e^{2\pi i s(u^1_0-u^2_0-i)}e^{2\pi i m_1 (u^1_0+u^2_0)}\,.
\end{split}
\end{equation}
where we used permutation $u^1_0 \leftrightarrow u^2_0$ and $s \to -s$ in the second term above.

Integration over any of the real variables, say $u^1_0$, can be written as an integration on the complex plane over a contour which goes along the real axis and closes in the upper-half plane. If one integrates the same function but over a contour shifted by unit distance in the imaginary direction compared to the previous contour (implemented by simply replacing $u^1_0 \to u^1_0 + i$ in the original integrand), the two integrals will differ by the sum of the residues that lie between $0<\text{Im}(u^1_0)<i$. Explicitly one gets
\begin{align}\label{Res1}
&\int  \frac{d^{2} u_{0}}{2!} ds \frac{e^{2 i \pi \kappa s^2}\sh{(u^1_{0} - u^2_{0})}}{\ch{s}\prod^2_{p,l=1} \ch{(u^p_{0} - u^l_{1})}}  e^{2\pi i s(u^1_0-u^2_0-i)}e^{2\pi i m_1 (u^1_0+u^2_0)}\cr
=&  \int \frac{d^{2} u_{0}}{2!} ds \frac{e^{2 i \pi \kappa s^2}\sh{(u^1_{0} - u^2_{0}+i)}}{\ch{s}\prod^2_{l=1} \ch{(u^1_{0}+i - u^l_{1})}\ch{(u^2_{0} - u^l_{1})}}  e^{2\pi i s(u^1_0+i-u^2_0-i)}e^{2\pi i m_1 (u^1_0+i+u^2_0)}\cr
&+ 2\pi i \sum_{l=1,2} \mbox{Res}_{u^1_0=u^l_1+i/2} \int  \frac{d^2 u_{0}}{2!} ds \frac{e^{2 i \pi \kappa s^2}\sh{(u^1_{0} - u^2_{0})}}{\ch{s}\prod^2_{p,l=1} \ch{(u^p_{0} - u^l_{1})}}  e^{2\pi i s(u^1_0-u^2_0-i)}e^{2\pi i m_1 (u^1_0+u^2_0)}\cr
=& - \int  \frac{d^{2} u_{0}}{2!} ds \frac{e^{2 i \pi \kappa s^2}\sh{(u^1_{0} - u^2_{0})}}{\ch{s}\prod^2_{p,l=1} \ch{(u^p_{0} - u^l_{1})}}  e^{2\pi i s(u^1_0-u^2_0)}e^{2\pi i m_1 (u^1_0+u^2_0+i)}\cr
&+ 2\pi i \sum_{l=1,2} \mbox{Res}_{u^1_0=u^l_1+i/2} \int  \frac{d^2 u_{0}}{2!} ds \frac{e^{2 i \pi \kappa s^2}\sh{(u^1_{0} - u^2_{0})}}{\ch{s}\prod^2_{p,l=1} \ch{(u^p_{0} - u^l_{1})}}  e^{2\pi i s(u^1_0-u^2_0-i)}e^{2\pi i m_1 (u^1_0+u^2_0)}.\notag \\
\end{align}
The integrand in the first term after the last equality is antisymmetric under the simultaneous operations $u^1_0 \leftrightarrow u^2_0$ and $s \to -s$ and therefore vanishes.
Now let us focus on the part depending on the residues:
\begin{align}
&2\pi i \sum_{l=1,2} \mbox{Res}_{u^1_0=u^l_1+i/2} \int  \frac{d^2 u_{0}}{2!} ds \frac{e^{2 i \pi \kappa s^2}\sh{(u^1_{0} - u^2_{0})}}{\ch{s}\prod^2_{p,l=1} \ch{(u^p_{0} - u^l_{1})}}  e^{2\pi i s(u^1_0-u^2_0-i)}e^{2\pi i m_1 (u^1_0+u^2_0)}\cr
=&-\int  d u^2_{0} ds \frac{e^{2 i \pi \kappa s^2}}{\ch{s}}\frac{1}{\sh{(u^1_1-u^2_1)}} \left(\frac{e^{i\pi s(-i+2u^2_1-2u^2_0)}e^{\pi i m_1 (2u^2_1+2u^2_0+i)}}{\ch{(u^2_0-u^1_1)}} - \frac{e^{i\pi s(-i+2u^1_1-2u^2_0)}e^{\pi i m_1 (2u^1_1+2u^2_0+i)}}{\ch{(u^2_0-u^2_1)}} \right)\cr
= & - \int  d u^2_{0} ds \frac{e^{2 i \pi \kappa s^2}}{\ch{s}} \frac{e^{\pi (s-m_1)}}{\sh{(u^1_1-u^2_1)}} \left(\frac{e^{2i\pi s(u^2_1+u^1_1-u^2_0)}e^{2\pi i m_1 (u^2_1-u^1_1+u^2_0)}}{\ch{(u^2_0-2 u^1_1)}} - \frac{e^{2i\pi s(u^1_1+u^2_1-u^2_0)}e^{2\pi i m_1 (u^1_1-u^2_1+u^2_0)}}{\ch{(u^2_0-2 u^2_1)}} \right)\cr
=& -\int  d u^2_{0} ds \frac{e^{2 i \pi \kappa s^2}}{\ch{s}} \times \frac{e^{\pi (s-m_1)} e^{2i\pi s(u^2_1+u^1_1-u^2_0)}e^{2\pi i m_1 u^2_0}}{\sh{(u^1_1-u^2_1)}}\cr
&\times \left(\frac{e^{2\pi i m_1 (u^2_1-u^1_1)}\ch{(u^2_0-2 u^2_1)}-e^{2\pi i m_1 (u^1_1-u^2_1)}\ch{(u^2_0-2 u^1_1)}}{\ch{(u^2_0-2 u^1_1)}\ch{(u^2_0-2 u^2_1)}} \right)\cr
=&-\int  d u^2_{0} ds \frac{e^{2 i \pi \kappa s^2}}{\ch{s}} \times \frac{e^{\pi (s-m_1)} e^{2i\pi s(u^2_1+u^1_1-u^2_0)}e^{2\pi i m_1 u^2_0}}{\sh{(u^1_1-u^2_1)}}\times \frac{1}{\ch{(u^2_0-2 u^1_1)}\ch{(u^2_0-2 u^2_1)}}\cr
&\times \left(\cos{2\pi m_1 (u^1_1-u^2_1)}\left((\ch{(u^2_0-2 u^2_1)}-\ch{(u^2_0-2 u^1_1)}\right) \right.\cr
&\left.- i \sin{2\pi m_1 (u^1_1-u^2_1)}(\ch{(u^2_0-2 u^2_1)}+\ch{(u^2_0-2 u^1_1)}) \right)\cr
\overset{m_1=0}=& -\int  d u^2_{0} ds \frac{e^{2 i \pi \kappa s^2}}{\ch{s}} \times e^{\pi s} e^{2i\pi s(u^2_1+u^1_1-u^2_0)}\left(\frac{2\sh{(u^2_0-u^2_1-u^1_1)}}{\ch{(u^2_0-2 u^1_1)}\ch{(u^2_0-2 u^2_1)}} \right)\cr
=& - \int  d u^2_{0} ds \frac{e^{2 i \pi \kappa s^2}}{\ch{s}} \times e^{\pi s} \left(\frac{e^{2i\pi (u^2_1+u^1_1-u^2_0)(s+i/2)} - e^{2i\pi (u^2_1+u^1_1-u^2_0)(s-i/2)}}{\ch{(u^2_0-2 u^1_1)}\ch{(u^2_0-2 u^2_1)}} \right).
\end{align}
A quick look at the fourth equality clearly suggests that a non-zero $m_1$ breaks the $U(2)$ gauge symmetry of the node associated with the double arrow. We set it to zero from here on\footnote{$m_1 \neq 0$ case does not seem to lead to an easy dual interpretation -- for example, it breaks the $U(2)$ gauge symmetry of the node parametrized by $\{u^l_1\}$}.

The integration over real variable $s$ can be written as an integration of a complex variable over a contour which goes along the real axis and closes in the upper-half (or lower-half) plane. As before, consider writing the above integral in terms of another integral with the same integrand but a contour that is shifted by a distance $-1/2$ in the imaginary direction, with any pole on the contour being traversed in an anti-clockwise fashion (just a convention -- nothing in the computation below depends on this choice). Therefore, the first term in the parentheses of the last equation may be rewritten as 
\begin{align}\label{eq:Zafirst}
&- \int  d u^2_{0} ds \frac{e^{2 i \pi \kappa s^2}}{\ch{s}} \times e^{\pi s} \left(\frac{e^{2i\pi (u^2_1+u^1_1-u^2_0)(s+i/2)}}{\ch{(u^2_0-2 u^1_1)}\ch{(u^2_0-2 u^2_1)}} \right)\cr
=& - \int  d u^2_{0} ds e^{2 i \pi \kappa (s-i/2)^2} \frac{1}{\ch{(s-i/2)}} e^{\pi(s-i/2)} \left(\frac{e^{2i\pi (u^2_1+u^1_1-u^2_0)(s-i/2+i/2)}}{\ch{(u^2_0-2 u^1_1)}\ch{(u^2_0-2 u^2_1)}} \right)\cr
&+ i \pi \mbox{Res}_{s=-i/2} \int d u^2_{0} \frac{e^{2 i \pi \kappa s^2}}{\ch{s}} \times e^{\pi s} \left(\frac{e^{2i\pi (u^2_1+u^1_1-u^2_0)(s+i/2)}}{\ch{(u^2_0-2 u^1_1)}\ch{(u^2_0-2 u^2_1)}} \right)\cr
=& - \int  d u^2_{0} ds e^{2 i \pi \kappa (s-i/2)^2} \frac{e^{\pi s}}{\sh{s}} \left(\frac{e^{2i\pi (u^2_1+u^1_1-u^2_0)s}}{\ch{(u^2_0-2 u^1_1)}\ch{(u^2_0-2 u^2_1)}} \right)\cr
&+ i \pi \mbox{Res}_{s=-i/2} \int d u^2_{0} \frac{e^{2 i \pi \kappa s^2}}{\ch{s}} \times e^{\pi s} \left(\frac{e^{2i\pi (u^2_1+u^1_1-u^2_0)(s+i/2)}}{\ch{(u^2_0-2 u^1_1)}\ch{(u^2_0-2 u^2_1)}} \right)\cr
=& - \int  d u^2_{0} ds e^{2 i \pi \kappa (s-i/2)^2} \frac{e^{\pi s}}{\sh{s}} \left(\frac{e^{2i\pi (u^2_1+u^1_1-u^2_0)s}}{\ch{(u^2_0-2 u^1_1)}\ch{(u^2_0-2 u^2_1)}} \right)\cr
&+ i {e^{-i\kappa \pi/2}} \int d u^2_{0}  \left(\frac{1}{\ch{(u^2_0-2 u^1_1)}\ch{(u^2_0-2 u^2_1)}} \right)\,.
\end{align}

Similarly, the second term can be written as
\begin{align}\label{eq:Zasecond}
& \int  d u^2_{0} ds \frac{e^{2 i \pi \kappa s^2}}{\ch{s}} \times e^{\pi s} \left(\frac{e^{2i\pi (u^2_1+u^1_1-u^2_0)(s-i/2)}}{\ch{(u^2_0-2 u^1_1)}\ch{(u^2_0-2 u^2_1)}} \right)\cr
=& \int  d u^2_{0} ds e^{2 i \pi \kappa (s+i/2)^2} \frac{1}{\ch{(s+i/2)}} e^{\pi(s+i/2)} \left(\frac{e^{2i\pi (u^2_1+u^1_1-u^2_0)(s+i/2-i/2)}}{\ch{(u^2_0-2 u^1_1)}\ch{(u^2_0-2 u^2_1)}} \right)\cr
&+i \pi \mbox{Res}_{s=i/2} \int d u^2_{0} \frac{e^{2 i \pi \kappa s^2}}{\ch{s}} \times e^{\pi s} \left(\frac{e^{2i\pi (u^2_1+u^1_1-u^2_0)(s-i/2)}}{\ch{(u^2_0-2 u^1_1)}\ch{(u^2_0-2 u^2_1)}} \right)\cr
=&\int  d u^2_{0} ds e^{2 i \pi \kappa (s+i/2)^2} \frac{e^{\pi s}}{\sh{s}} \left(\frac{e^{2i\pi (u^2_1+u^1_1-u^2_0)s}}{\ch{(u^2_0-2 u^1_1)}\ch{(u^2_0-2 u^2_1)}} \right)\cr
&+ i {e^{-i\kappa \pi/2}} \int d u^2_{0}  \left(\frac{1}{\ch{(u^2_0-2 u^1_1)}\ch{(u^2_0-2 u^2_1)}} \right).
\end{align}

Therefore, we find after adding \eqref{eq:Zafirst} and \eqref{eq:Zasecond}
\begin{align}
&\frac{1}{2}\int  \frac{d^{2} u_{0}}{2!} ds \frac{e^{2 i \pi \kappa s^2}\sh{(u^1_{0} - u^2_{0})}}{\ch{s}\prod^2_{p,l=1} \ch{(u^p_{0} - u^l_{1})}}  e^{2\pi i s(u^1_0-u^2_0-i)}\nonumber\\
=& - e^{-i\kappa\pi/2} \int  d u^2_{0} ds e^{2 i \pi \kappa s^2} \frac{e^{\pi s}}{\sh{s}}\sh{2
\kappa s}\left(\frac{e^{2i\pi (u^2_1+u^1_1-u^2_0)s}}{\ch{(u^2_0-2 u^1_1)}\ch{(u^2_0-2 u^2_1)}} \right)\nonumber \\
& + i {e^{-i\kappa \pi/2}} \int d u^2_{0}  \left(\frac{1}{\ch{(u^2_0-2 u^1_1)}\ch{(u^2_0-2 u^2_1)}} \right),\\
\overset{\kappa=1/2}
=&- \int  d u^2_{0} ds e^{i \pi (s-i/2)^2}  \left(\frac{e^{2i\pi (u^2_1+u^1_1-u^2_0)s}}{\ch{(u^2_0-2 u^1_1)}\ch{(u^2_0-2 u^2_1)}} \right)\nonumber \\
& + i {e^{-i\pi/4}} \int d u^2_{0}  \left(\frac{1}{\ch{(u^2_0-2 u^1_1)}\ch{(u^2_0-2 u^2_1)}} \right)\\
\overset{s \to s+i/2}
=& - {e^{i\pi/4}}\int  d u^2_{0}  e^{-i \pi (u^2_1+u^1_1-u^2_0)^2}  \left(\frac{e^{-\pi (u^2_1+u^1_1-u^2_0)}}{\ch{(u^2_0-2 u^1_1)}\ch{(u^2_0-2 u^2_1)}} \right)\nonumber \\
& +i {e^{-i\pi/4}} \int d u^2_{0}  \left(\frac{1}{\ch{(u^2_0-2 u^1_1)}\ch{(u^2_0-2 u^2_1)}} \right)\\
\overset{u^2_0 \to u^2_0+u^2_1+u^1_1}
=& - {e^{i\pi/4}}\int  d u^2_{0}  e^{-i \pi (u^2_0)^2}  \left(\frac{e^{\pi u^2_0}}{\ch{(u^2_0-u^1_1+u^2_1)}\ch{(u^2_0- u^2_1+u^1_1)}} \right)\nonumber \\
& + i {e^{-i\pi/4}} \int d u^2_{0}  \left(\frac{1}{\ch{(u^2_0-2 u^1_1)}\ch{(u^2_0-2 u^2_1)}} \right)\,,
\end{align}
which leads us to \eqref{nsl0}.

\section{Generalities on partition functions and superconformal indices on $S^3/\mathbb{Z}_r \times S^1$}\label{Sec:GenIndex}
In this section we list the rules for deforming a partition function on $S^3$ to the 4d index, which can be thought of as partition function on $S^3\times S^1$. The integration variables $s^{(\beta)} (\beta=0,1,3), s^{(\beta)}_i$ lie in the Cartan subalgebra of the gauge group $U(1)^3 \times U(2)$ corresponding to the framed affine $B_{3}$ quiver theory (see \figref{fig:quiverbN})

In order to write the index, we define corresponding fugacities as $z^{(\beta)}= e^{2\pi i s^{(\beta)}}$, $z^{(2)}_i=e^{2\pi i s^{(\beta)}_i }$. Recall that the superconformal index for a $4d$, $\cN=2$ theory on lens space $L(1,r)$ is defined as
\begin{equation}\label{4dindex-def1}
\mathcal{I}(p,q,t;z_i)= \tr_{S^3/\mathbb{Z}_r} \left[(-1)^F \left(\frac{t}{pq}\right)^{R'} p^{j_2+j_1} q^{j_2-j_1} t^R  e^{-\beta(E-2j_2 -2R +R')} \prod_i z_i^{f_i}\right]
\end{equation}
where the trace is taken over the Hilbert space on $S^3/\mathbb{Z}_r$, $F$ denotes the fermion number, $j_1$, $j_2$ the Cartans 
of the rotation group $SU(2)_1 \times SU(2)_2 \sim SO(4)$, $R$ the $U(1)$ generator of $SU(2)_R$ R-symmetry and $R'$ the generator of $U(1)_R$, and $f_i$ possible flavor symmetries (some of which may be gauged).\\

A crucial difference between the Lens space index and the $S^3 \times S^1$ index is that in the former case one can turn on non-trivial discrete holonomies along the Hopf fiber of the Lens space for the gauge (flavor) vector fields -- parametrized by integers $m^{(\alpha)}_i$ $(\widetilde{m}^{(\kappa)}_i)$ for every gauge (flavor) node $\alpha$ ($\kappa$) where  $0 \leq m^{(\alpha)} _i< r$.  For a simply-connected group $G$ (gauge or flavor), the discrete holonomy $V$ of the vector field  may be represented as elements in the Cartan of the group $G$: $V=\text{diag}(e^{2\pi i m_1/r}, \dots,e^{2\pi i m_N/r})$ where $N=\text{rank}(G)$. The 4d index therefore involves a sum over these integers $\{m_i^{(\alpha)}\}$.\\

 In terms of indices of $\cN=2$ vector multiplet and hypermultiplet, the index of a quiver gauge theory with gauge group $G=\prod_\alpha U(N_\alpha)$ and bifundamental and fundamental matter may be written as 
\begin{equation}
\begin{split}
&\mathcal{I}\left(p,q,t;\widetilde{z}^{(\alpha)}, \widetilde{m}^{(\alpha)}\right)=\\ 
& \sum_{m^{(\alpha)}} \oint\limits_{|z_i|=1} \prod_{\alpha} \frac{1}{W_\alpha(m^{(\alpha)})} \prod^{N_\alpha}_{i=1} \frac{dz^{(\alpha)}_i}{2\pi i z^{(\alpha)}_i}\mathcal{I}_V^{(m^{(\alpha)})}(z^{(\alpha)}) \mathcal{I}_\text{fund}^{(m^{(\alpha)}, \widetilde{m}^{(\alpha)})} (z^{(\alpha)}, \widetilde{z}^{(\alpha)})\prod_{(\beta,\gamma)} \mathcal{I}_\text{bifund}^{(m^{(\beta)}, m^{(\gamma)})} (z^{(\beta)}, z^{(\gamma)}) .
\end{split}
\end{equation}
where $\{\widetilde{z}^{(\alpha)} , \widetilde{m}^{(\alpha)} \}$ denote respectively fugacities and discrete holonomies of the flavor node $\alpha$ in the quiver diagram. The individual factors in the integrand may be identified as follows:\\
\\
$\mathcal{I}_V^{(m^{(\alpha)})}(z^{(\alpha)})\equiv$ index of the vector multiplet corresponding to the $\alpha$-th gauge node in the quiver diagram and $\alpha$ runs over all gauge nodes in the quiver. \\
$\mathcal{I}_\text{bifund}^{(m^{(\beta)}, m^{(\gamma)})} (z^{(\beta)}, z^{(\gamma)})\equiv$ index of a bifundamental hyper and $(\beta,\gamma)$ runs over all lines connecting two gauge nodes in the quiver.\\
$\mathcal{I}_\text{fund}^{(m^{(\alpha)}, \widetilde{m}^{(\alpha)})} (z^{(\alpha)}, \widetilde{z}^{(\alpha)})\equiv$ index of a fundamental hyper at the gauge node $\alpha$ and $\alpha$ runs over all gauge nodes in the quiver.\\
\\
Note that in the above formula we have cancelled the Haar measure of the integral against a similar factor coming from contributions of the vector multiplets to the index. Accounting for this overall factor, the explicit form for the vector multiplet index is given in terms of elliptic gamma function $\Gamma(z;p,q)=\prod^{\infty}_{i,j \geq 0} \frac{1-p^{i+1} q^{j+1} z^{-1}}{1- p^i q^j z}$ and the $q$-Pochammer symbol $(z;q)=\prod^\infty_{l=0} (1-z q^l)$ as follows:
\begin{equation}
\begin{split}
&\mathcal{I}_V^{(m^{(\alpha)})}(z^{(\alpha)})= \left(\frac{(p^r;p^r)}{\Gamma(t;pq,p^r)} \frac{q^r;q^r}{\Gamma(tq^r;pq,q^r)}\right)^{N_\alpha} \prod_{\rho \in Adj^{(\alpha)}}\left(\frac{pq}{t}\right)^{-\frac{1}{2}([[\rho(m^{(\alpha)})]]-\frac{1}{r}[[\rho(m^{(\alpha)})]]^2 )}\\
& \times \frac{1}{\Gamma(tp^{[[\rho(m^{(\alpha)})]]} e^{2\pi i \rho(s^{(\alpha)})};pq,p^r)} \frac{1}{\Gamma(tq^{r-[[\rho(m^{(\alpha)})]]} e^{2\pi i \rho(s^{(\alpha)})};pq,p^r)}\\
& \times \frac{1}{\Gamma(p^{[[\rho(m^{(\alpha)})]]} e^{2\pi i \rho(s^{(\alpha)})};pq,p^r)} \frac{1}{\Gamma(q^{r-[[\rho(m^{(\alpha)})]]} e^{2\pi i \rho(s^{(\alpha)})};pq,p^r)}.
\end{split}
\end{equation}
where $s^{(\alpha)}_i$ lies in the Cartan subalgebra of the gauge group at the node $\alpha$ with $z^{(\alpha)}_i=e^{2\pi i s^{(\alpha)}_i}$ and $[[x]]$ is defined as $x=[[x]]$ modulo $r$. The product is over all roots of the Lie algebra of the gauge group. For an Abelian gauge theory, the contribution of the vector multiplet index is trivial.\\

The contributions of the bifundamental and fundamental hyper are given as
\begin{equation}
\begin{split}
&\mathcal{I}_\text{bifund}^{(m^{(\beta)}, m^{(\gamma)})} (z^{(\beta)}, z^{(\gamma)})=\prod_{s=\pm 1}\prod_{\rho \in Bif^{(\beta,\gamma)}}\left(\frac{pq}{t}\right)^{\frac{1}{4}([[s\rho(m^{(\beta)}, m^{(\gamma)})]]-\frac{1}{r}[[s\rho(m^{(\beta)}, m^{(\gamma)})]]^2 )}\\
& \times \prod_{s=\pm 1} \Gamma(t^{1/2} p^{[[s\rho(m^{(\beta)}, m^{(\gamma)})]]} e^{2\pi i s \rho(s^{(\beta)},s^{(\gamma)})};pq,p^r) \Gamma(t^{1/2} q^{r-[[s\rho(m^{(\beta)}, m^{(\gamma)})]]}e^{2\pi i s \rho(s^{(\beta)},s^{(\gamma)})} ;pq,q^r), \\
&\mathcal{I}_\text{fund}^{(m^{(\alpha)}, \widetilde{m}^{(\alpha)})} (z^{(\alpha)}, \widetilde{z}^{(\alpha)})=\prod_{s=\pm 1} \prod_{\rho \in Bif^{(\alpha,\alpha)}}\left(\frac{pq}{t}\right)^{\frac{1}{4}([[s\rho(m^{(\alpha)}, \widetilde{m}^{(\alpha)})]]-\frac{1}{r}[[s\rho(m^{(\alpha)}, \widetilde{m}^{(\alpha)})]]^2 )}\\
& \times \prod_{s=\pm 1} \Gamma(t^{1/2} p^{[[s\rho(m^{(\alpha)}, \widetilde{m}^{(\alpha)})]]} e^{2\pi i s \rho(s^{(\alpha)}, \widetilde{s}^{(\alpha)})};pq,p^r) \Gamma(t^{1/2} q^{r-[[s\rho(m^{(\alpha)}, \widetilde{m}^{(\alpha)})]]}e^{2\pi i s \rho(s^{(\alpha)}, \widetilde{s}^{(\alpha)})} ;pq,q^r).
\end{split}
\end{equation}
For generic matter in some representation $R$, the formula for the index is exactly the same with $\rho$ now being a weight of the representation $R$.

\section{Folding}\label{Sec:Folding}
\begin{figure}[!h]
\begin{center}
\includegraphics[scale=0.4]{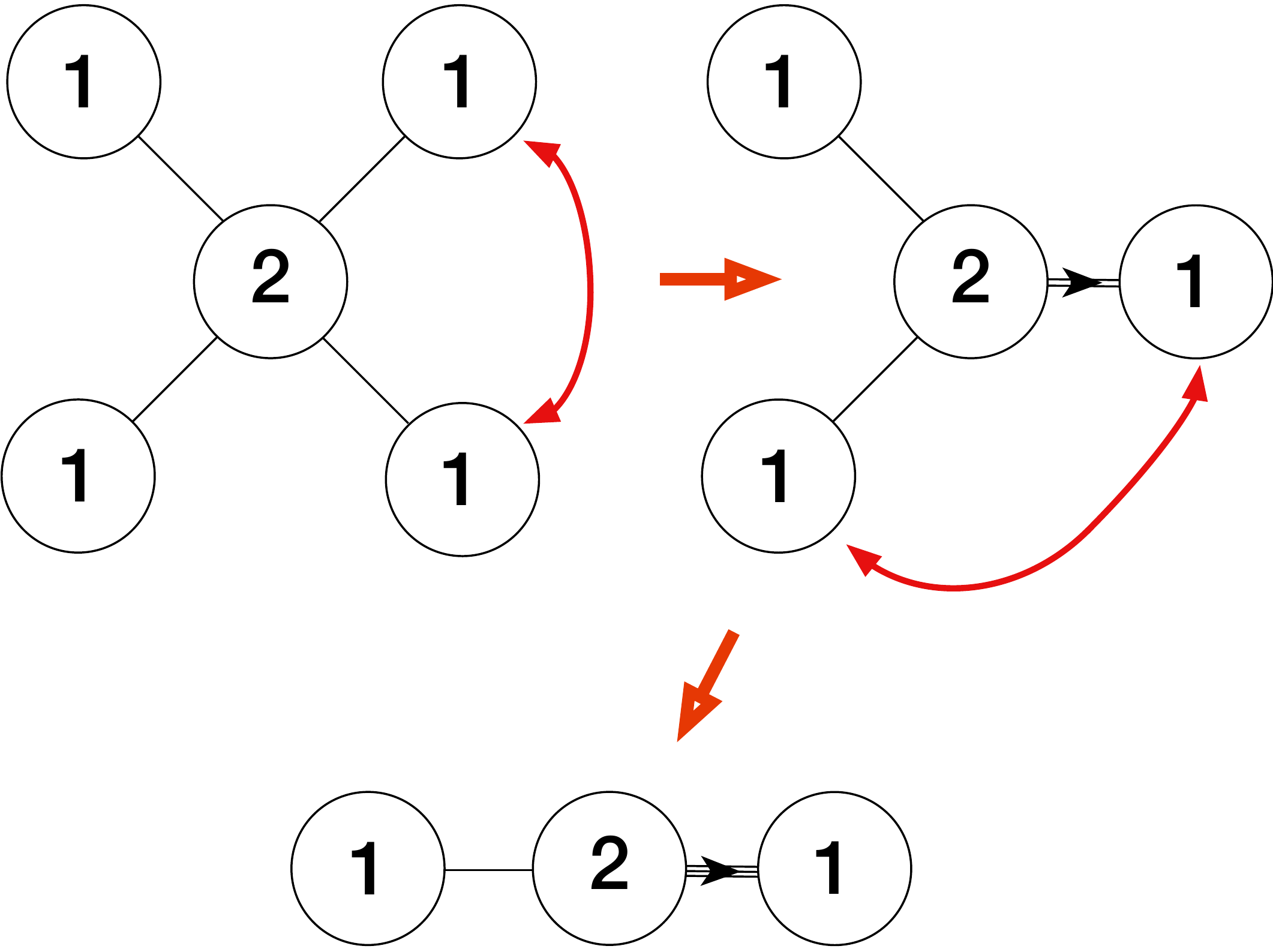}
\caption{Folding $D_4$ Dynkin diagram to $B_3$ Dynkin diagram, and then to $G_2$ Dynkin diagram.}
\label{fig:D4B3G2}
\end{center}
\end{figure}

Folding is a standard operation of converting ADE-type Dynkin graphs into other types of Dynkin graphs \cite{2005math.....10216S}. In the context of four dimensional theories of class $\mathcal{S}$ folding was discussed in \cite{Cecotti:2012fj}. Seiberg-Witten theories with $Spin(2N-1)$ groups were obtained from $Spin(2N)$ theories in \cite{Chacaltana:2014aa} by a similar mechanism which is discussed later in this section. An example of folding is depicted in 
\figref{fig:D4B3G2}

In physics context folding of Dynkin diagrams has already been discussed in the literature. 
In \cite{Hanany:2012aa} the authors computed Higgs branch Hilbert series for 3d $\CN=4$ quiver theories, which describe moduli space of instantons of BCFG types, by exploiting the folding technique in order to obtain non-simply laced quivers from ADE type quivers, for which the computation was known. Later in \cite{Cremonesi:2014xha} it was shown how to compute Coulomb branch Hilbert series for the moduli space of $G$-instantons for any simple Lie group $G$. 

Therefore it is not known how to describe both Higgs and Coulomb branches of the ADHM quiver theories and their mirror duals, e.g. \figref{fig:D4S08Mirror}. Therefore using the results of \cite{Hanany:2012aa, Cremonesi:2014xha} and some other developments we can study physics of the non-simply laced quiver gauge theories (like the left quiver in \figref{fig:D4S08Mirror}) which feature double and triple arrows\footnote{If we include affine and twisted affine series then quadruple arrows may also appear.}. In particular we should be able to understand what kind of matter fields correspond to those multiple arrows on the diagram. Also, by using folding technique, we will be able to realized those fields via gauging of \textit{discrete} global symmetries of the original quiver theories. These problems will be addressed in the future publications, however, in the end of this paper we shall discuss some ideas which should be further developed.

\subsection{Classical Analysis}
In addition we can analyze the dual theories in \figref{fig:B4S07Mirror} by studying their parameter spaces of supersymmetric vacua along the lines of \cite{Dey:2014jk}. The quiver gauge theory is studied on a cylinder $\mathbb{R}^2\times S^1_R$ of radius $R$ in the presence of the $\CN=2^*$ mass deformation parameter $\epsilon$. After the mass deformation the Coulomb branch of the theory degenerates into a set of discrete massive vacua whose position is determined by the twisted F-term relations which now depend on the $\CN=2^*$ mass $\eta=e^{Rm}$ (see \cite{Gaiotto:2013qy} for details). Below we shall analyze the corresponding twisted F-term relations\footnote{In gauge/integrability correspondence \cite{Nekrasov:2009uh} they coincide with Bethe Ansatz equations for an exactly soluble lattice model.} for $\widehat{D}_4$ and its folded version $\widehat{B}_3$. 

\begin{figure}[t]
\begin{center}
\includegraphics[scale=.4]{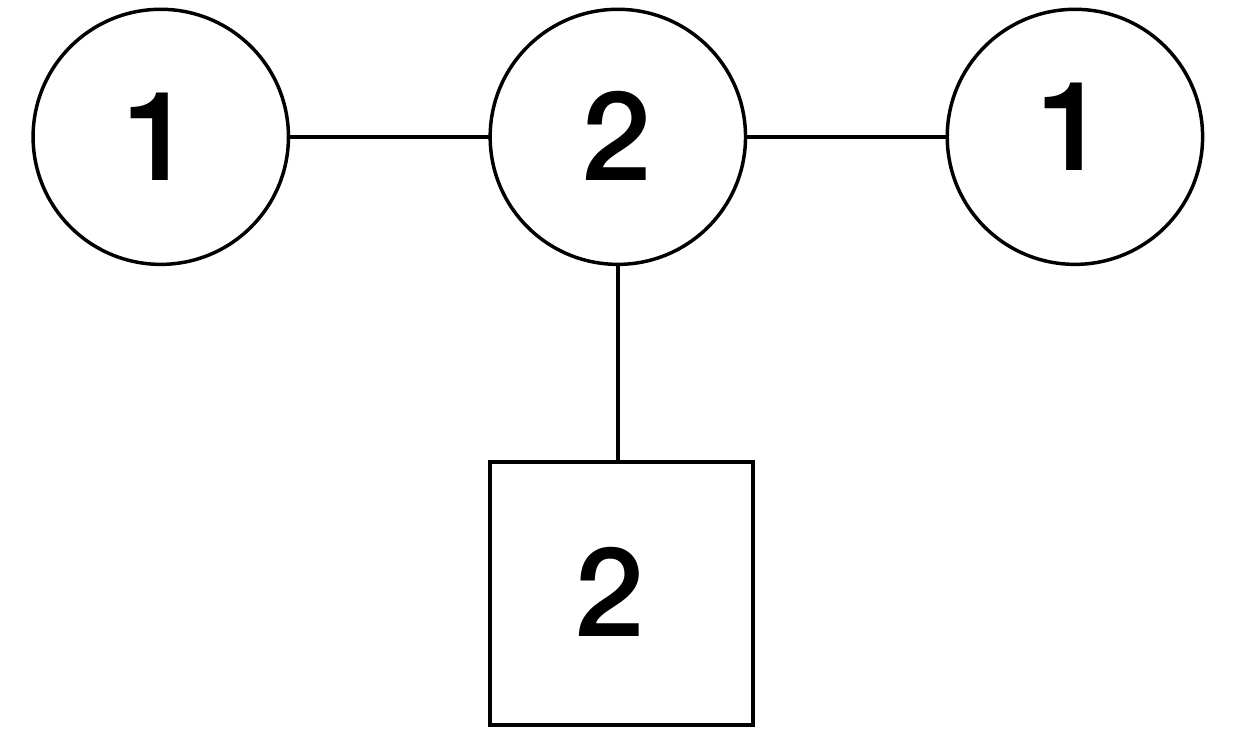} \qquad \qquad \includegraphics[scale=.4]{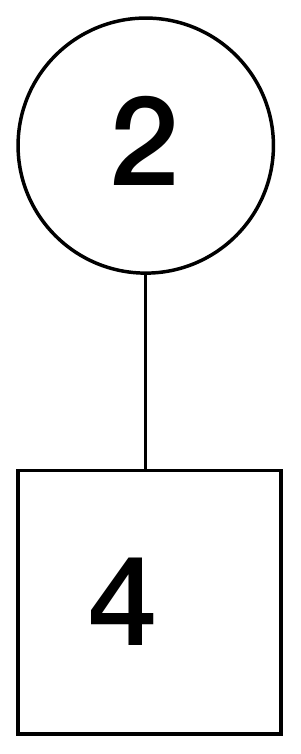}
\caption{Mirror dual $A_3$ and $A_1$ quivers with framing.}
\label{Fig:A3A1mirror}
\end{center}
\end{figure}

\begin{figure}[!h]
\begin{center}
\includegraphics[scale=0.5]{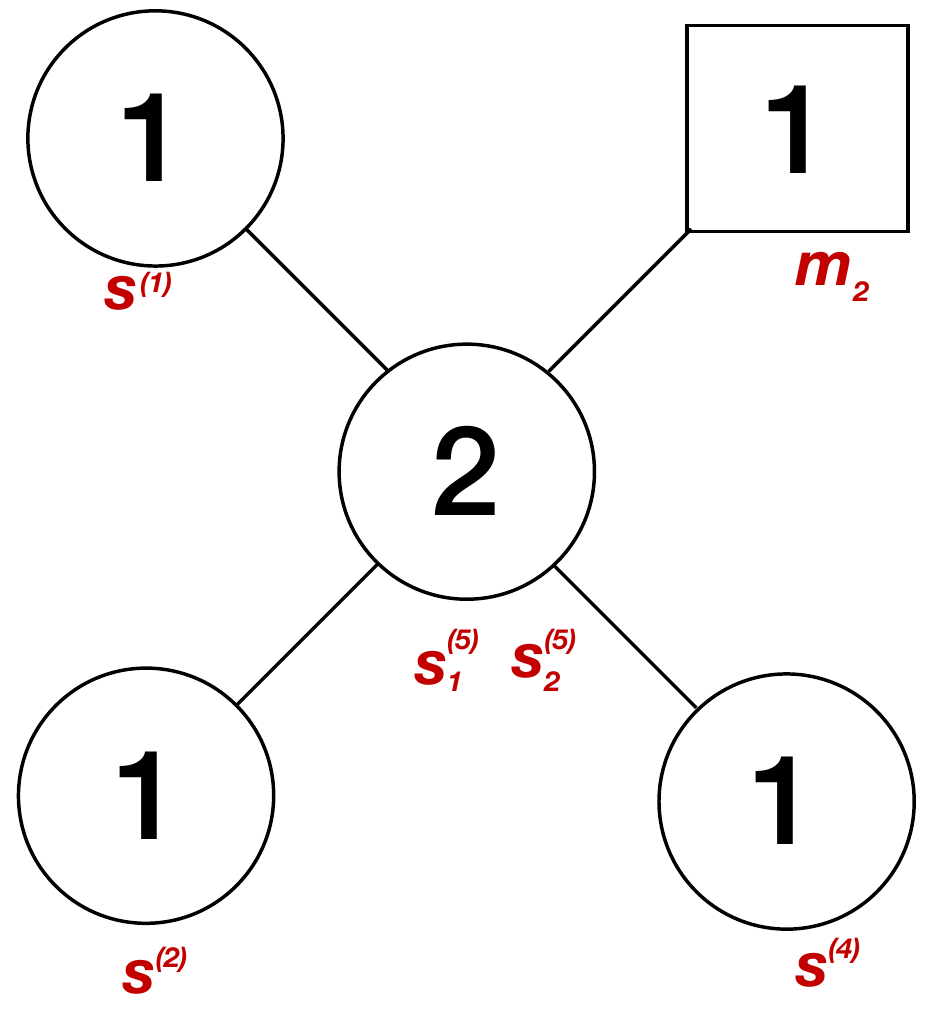}
\caption{$\widehat D_4$ quiver with labels.}
\label{fig:quiverD4hatIns}
\end{center}
\end{figure}

It was shown in \cite{Dey:2014jk} that both theories in \figref{fig:D4S08Mirror} can be obtained by gauging and ungauging global symmetries in the mirror pair represented by two A-type quivers with framing depicted in \figref{Fig:A3A1mirror}.
After gauging a $U(1)\subset U(2)$ global symmetry for the theory on the left one obtains a $\widehat{D}_4$-shaped quiver as shown in \figref{fig:quiverD4hatIns}. 
Its mirror is the $Sp(1)$ theory with $SO(8)$ global symmetry. For the latter we can write (see \cite{Dey:2014jk})
\begin{equation}
\mu_2\prod_{i=1}^3\frac{\eta^{-1}\sigma-\tau_i}{\eta^{-1}\tau_i-\sigma}\cdot \frac{\eta\sigma-\eta^{-1}/\sigma}{\eta\sigma-\eta^{-1}/\sigma}\cdot\frac{\eta^{-1}\sigma-\tau_4}{\eta^{-1}\tau_4-\sigma}=1\,,
\end{equation}
where we have singled out the contribution from the twisted hypermultiplet with mass $\tau_4$ in the last term. It is also required that $\tau_1^2=1$. The canonical momenta are 
\begin{equation}
p_\tau^{2\,\vee} = \tau_1\tau_2\tau_3\tau_4\,, \quad p_\mu^{a\,\vee}=\mu_2\prod_{i=1}^2\frac{\eta^{-1}\tau_a+\sigma_i}{\eta^{-1}\sigma_i+\tau_a}\,.
\end{equation}
Let us focus on the last term in the above equation. We implement the following scaling
\begin{equation}
\tau_4\to x\tau_4\,,\quad \widetilde\eta\to x^{-1}\widetilde\eta\,, \quad x\to\infty\,,
\label{eq:scalingB}
\end{equation}
where we have substituted $\eta$ with $\widetilde\eta$. Then the last term above becomes
\begin{equation}
\mu_2\frac{\sigma-\widetilde\eta\tau_4}{\tau_4-\widetilde\eta\sigma}\to \frac{\mu_2}{\tau_4}(\sigma-\widetilde\eta\tau_4)
\end{equation}
if in addition we scale $\mu_2\to x\mu_2$.

For the A model we have 
\begin{align}
&\frac{\tau_4\tau_3}{p_\mu^2}\prod_{i=1}^2\frac{\eta\mu_2-\sigma_i^{(5)}}{\eta\sigma_i^{(5)}-\mu_2}=1\,,\quad \frac{\tau_4}{\tau_3}\prod_{i=1}^2\frac{\eta\sigma^{(4)}-\sigma_i^{(5)}}{\eta\sigma_i^{(4)}-\sigma^{(3)}}=1\,,\notag\\
&\frac{\tau_3}{\tau_2}\prod_{I=1}^2\frac{\eta\sigma^{(5)}_i-\sigma^{(I)}}{\eta\sigma^{(I)}-\sigma^{(5)}_i}\prod_{j\neq i}\frac{\eta^{-1}\sigma^{(5)}_i-\eta\sigma_j^{(5)}}{\eta^{-1}\sigma^{(5)}_j-\eta\sigma_i^{(5)}}\cdot\frac{\eta\sigma^{(5)}_i-\mu_2}{\eta\mu_2-\sigma^{(5)}_i}\frac{\eta\sigma^{(5)}_i-\sigma^{(4)}}{\eta\sigma^{(4)}-\sigma^{(5)}_i}=1\,.
\label{eq:AmodelBetheEq}
\end{align}  
together with the momenta
\begin{equation}
p_\tau^4 = \mu_2\sigma^{(4)}\,,\quad p_\tau^3 = \mu_2\frac{\sigma^{(5)}_1\sigma^{(5)}_2}{\sigma^{(4)}}\,,
\end{equation}
as well as $p_\mu^1 = \tau_3^2$ as is required by gauging. Now we need to implement scaling \eqref{eq:scalingB} together with $\mu_2\to\infty$ as before using $\widetilde\eta$ instead of $\eta$ for the $\sigma^{(4)}$ node. One has from \eqref{eq:AmodelBetheEq}
\begin{align}
&\frac{\tau_4\tau_3}{p_\mu^2}\prod_{i=1}^2\frac{\widetilde\eta\mu_2-\sigma_i^{(5)}}{-\mu_2}=1\,,\quad \frac{\tau_4}{\tau_3}\prod_{i=1}^2\frac{\widetilde\eta\sigma^{(4)}-\sigma_i^{(5)}}{-\sigma^{(4)}}=1\,,\notag\\
&\frac{\tau_3}{\tau_2}\prod_{I=1}^2\frac{\eta\sigma^{(5)}_i-\sigma^{(I)}}{\eta\sigma^{(I)}-\sigma^{(5)}_i}\prod_{j\neq i}\frac{\eta^{-1}\sigma^{(5)}_i-\eta\sigma_j^{(5)}}{\eta^{-1}\sigma^{(5)}_j-\eta\sigma_i^{(5)}}\cdot\frac{-x\mu_2}{\widetilde\eta\mu_2-\sigma^{(5)}_i}\frac{-x\sigma^{(4)}}{\widetilde\eta\sigma^{(4)}-\sigma^{(5)}_i}=1\,.
\label{eq:AmodelBetheEqScaled}
\end{align}  
We have implemented some additional scaling
\begin{equation}
\sigma^{(4)} \to x\sigma^{(4)}
\end{equation}
Finally we gauge the remaining global $U(1)$ by setting similar to \cite{Dey:2014jk}
\begin{equation}
\frac{\tau_4\tau_3}{p_\mu^2}=\frac{\tau_4}{\tau_3}\,,
\end{equation}
so the first and the second equations of \eqref{eq:AmodelBetheEqScaled} become the same and one identifies $\mu_2=\sigma^{(4)}$.  Therefore the Bethe equation for the middle node reads
\begin{equation}
\frac{\tau_3}{\tau_2}\prod_{I=1}^2\frac{\eta\sigma^{(5)}_i-\sigma^{(I)}}{\eta\sigma^{(I)}-\sigma^{(5)}_i}\prod_{j\neq i}\frac{\eta^{-1}\sigma^{(5)}_i-\eta\sigma_j^{(5)}}{\eta^{-1}\sigma^{(5)}_j-\eta\sigma_i^{(5)}}\cdot\left(\frac{\sigma^{(4)}}{\widetilde\eta\sigma^{(4)}-\sigma^{(5)}_i}\right)^2=1\,.
\end{equation}
We can recognize the contribution from the double arrow in the last term which is a square of a rational function. One can clearly see that this contribution cannot be reproduced by integrating out any (bi)fundamental matter, thus it represents a new contribution, which is certainly non-Lagrangian.

\subsection{Chern-Simons terms for the ADHM quiver}
In the example in \secref{Sec:Intro} we compared dimensions of Higgs and Coulomb branches of the ADHM quivers with $SO(8)$ and $SO(7)$
global symmetry. Here we shall remind the reader that if one integrates out a single half-hypermultiplet (e.g. to arrive to $SO(7)$ flavor group starting from $SO(8)$) the Chern-Simons term with level $1/2$ gets generated.

Let us start with the partition function of 3d $\CN=4$ $SU(2)$ gauge theory with $SO(8)$ symmetry on a squashed three-sphere \cite{Hama:2011ea} with squashing parameter $b$
\begin{equation}
\CZ_{S^3_b}=-8\int ds \sinh (2\pi i b^\pm s) S(\varepsilon+2s) \cdot \prod\limits_{a=1}^4 S\left(\frac{\varepsilon}{2}\pm(\pm s-m_a)\right)\,,
\label{eq:Z3bSU2SO8}
\end{equation}
where the integration is performed along the real $s$ line. 
The integrand consists of the vector multiplet contribution followed by the product of eight half-hypers. Here $2\varepsilon=b+b^{-1}$ and $\pm$ signs in the integrand show that the product is taken over all possible sign choices. Thus there are sixteen $S(z)$ functions overall in the half-hyper contribution.

In order to reduce the global symmetry to $SO(7)$ we can gauge discrete $\mathbb{Z}_2$ symmetry from the Weyl group of $SO(8)$ by integrating out one of the eight half-hypers. There are four terms involving $m_4$ in \eqref{eq:Z3bSU2SO8}. Gauging of $\mathbb{Z}_2$ symmetry will consist from two steps. First we break the $\mathbb{Z}_2$ symmetry by introducing a new mass parameter for two of the above four terms 
\begin{equation}
S\left(\frac{\varepsilon}{2}\pm(s-m_4)\right)S\left(\frac{\varepsilon}{2}\pm(-s-\widetilde m_4)\right)\,.
\end{equation}
Second, we integrate over $\widetilde m_4$. 
Recall that at large values of the argument the double sine function has the following behavior
\begin{equation}
S(z)\sim e^{\frac{\pi i}{2}B_{2,2}(z)}\,,
\end{equation}
where $B_{2,2}(z)=z^2+\varepsilon z + \frac{b^2+b^{-2}+3}{6}$. The latter constant will not be important for our analysis. Given the above asymptotic we have
\begin{equation}
S\left(\frac{\varepsilon}{2}\pm(-s-\widetilde m_4)\right) \sim e^{\frac{i\pi}{4}\left(4(\widetilde m_4)^2+4s^2+\varepsilon^2\right)}
\end{equation}
A trivial Gaussian integration gives the desired $SU(2)$ Chern-Simons term with level $\kappa=1/2$
\begin{equation}
\label{eq:CStermGauss}
\mathcal{Z}_\text{CS}\sim e^{i\pi s^2}\,.
\end{equation}

\section{Hilbert series}\label{Sec:HSApp}
\subsection{Coulomb Branch Hilbert series}
We can use the Coulomb branch monopole formula \cite{Cremonesi:2014xha} to write the Hilbert series for the $\widehat{D}_4$ quiver in \figref{fig:quiverD4hatIns} and study the folding trick. On the mirror side we may use the Higgs branch formula to understand how the the global $SO(8)$ symmetry is reduced down to $SO(7)$.

Let us first look at the Coulomb branch of the $\widehat{D}_4$. Scaling dimensions of monopole operators of quiver from \figref{fig:quiverD4hatIns} read
\begin{equation}
2\Delta_8 = \sum_{i=1}^3\sum_{j=5,6} |m_i-m_j| - 2|m_5-m_6|\,.
\label{eq:Delta8MonDim}
\end{equation}
After the folding is done we need to identify two nodes, in this case they are nodes 3 and 4 we identify 
\begin{equation}
m_3\to \frac{m_3}{2}\,\quad m_4\to \frac{m_3}{2}\,.
\label{eq:monopolechargemapping}
\end{equation}
The monopole formula then reads
\begin{equation}
2\Delta_7 = \sum_{i=1}^2\sum_{j=5,6} |m_i-m_j| +\sum_{j=5,6}|m_3-2m_j| - 2|m_5-m_6|\,.
\label{eq:monopoleformulalongroot}
\end{equation}
The Coulomb branch Hilbert series for the $\widehat{D}_4$ quiver reads \cite{Cremonesi:2013lqa}
\begin{equation}
H(t,z_1,z_2,z_3,z_4) = \sum_{m_1,\dots, m_6} t^{\Delta_8} P(t,m_1,\dots m_6) z_1^{m_1} z_2^{m_5+m_6} z_3^{m_3} z_4^{m_4}\,,
\end{equation}
where $\Delta_8$ is given by \eqref{eq:Delta8MonDim}. The Hilbert series can be thought of as a sum over the root lattice of the Lie algebra weighted by the scaling dimension of the monopole operators $\Delta$. The contribution with the lowest value $\Delta=1$ contains the following terms
\begin{equation}
z_1, z_2, z_3, z_4,\,\, z_1 z_2, z_3 z_2, z_4 z_2, \,\, z_1 z_2 z_3, z_1 z_2 z_4, z_4 z_2 z_3, \,\, z_1 z_2 z_3 z_4, \,\, z_1 z_2^2 z_3 z_4\,,
\label{eq:CouldombbranchGensDelta18}
\end{equation}
which correspond to twelve simple roots of $SO(8)$. We can manifestly see the $SO(8)$ triality which interchanges $z_1, z_2$ and $z_3$. 

Let us now apply the folding trick to the $\widehat{D}_4$ quiver, namely we apply \eqref{eq:monopolechargemapping} together with identifying $z_4$ with $z_3$. Then the above nine terms at $\Delta=1$ become
\begin{equation}
z_1, z_2, z_3,\,\, z_1 z_2, z_3 z_2, \,\, z_1 z_2 z_3, z_2 z^2_3, \,\, z_1 z_2 z^2_3,\,\, z_1 z_2^2 z^2_3\,,
\label{eq:CouldombbranchGensDelta1}
\end{equation}
which correspond to nine simple roots of $SO(7)$. Therefore we were able to verify the validity of the monopole formula \eqref{eq:monopoleformulalongroot} by folding.

\subsection{Higgs Branch Hilbert series}
On the mirror side we have $Sp(1)$ gauge theory with eight half-hypers. In order to understand the transition from $SO(8)$ global symmetry to $SO(7)$ global symmetry one half-hypermultiplet has to be removed which can be implemented by giving it a large mass. Let us verify that the number of the degrees of freedom after integrating out the half-hyper provides the correct matching with the Coulomb branch data given in \eqref{eq:CouldombbranchGensDelta1}. The global symmetry for the $SO(8)$ theory is parameterized by the $8\times 8$ antisymmetric matrix whose 28 nonzero components decompose as $28=4+12+12$ in terms of Cartan subalgebra generators, positive roots, and negative roots respectively. Indeed, \eqref{eq:CouldombbranchGensDelta18} contains 12 terms corresponding to the positive roots of $D_4$. After integrating out the half-hypermultiplet the 21 components  of the $7\times 7$ matrix decompose as $21=3+9+9$, again, in accordance with \eqref{eq:CouldombbranchGensDelta1}.

\bibliography{cpn1}
\bibliographystyle{JHEP}

\end{document}